%% file: main.tex
\PassOptionsToPackage{dvipsnames}{xcolor}
\documentclass[preprint]{aastex631}

\usepackage{amsmath}
\usepackage{bm}
\usepackage{capt-of}   
\usepackage{empheq}
\usepackage{enumitem}
\usepackage{float}
\usepackage{hyperref}
\usepackage{hhline}
\usepackage{longtable}
\usepackage{numprint}
\usepackage{pifont}
\usepackage{xspace}
\usepackage{placeins}
\usepackage{graphicx}
\usepackage{mwe}
\usepackage{multirow}
\usepackage{tikz}

\renewcommand{\noprint}[1]{}

\newcommand{\xh}{\textcolor{black}}
\newcommand{\sbj}{\textcolor{black}}

\newcommand{\ed}{\textcolor{black}}
\newcommand{\tblue}{\textcolor{black}}

\newcommand{\ho}{\ensuremath{H_0}\xspace}

\newcommand{\gigal}{\texttt{GIGA-Lens}\xspace}

\newcommand{\rhat}{\ensuremath{\hat{R}}\xspace}

\newcommand{\HST}{\emph{Hubble Space Telescope}\xspace}
\newcommand{\hst}{\emph{HST}\xspace}

\newcommand{\Ia}{SN~Ia\xspace}
\newcommand{\Iae}{SNe~Ia\xspace}
\newcommand{\snz}{SN~Zwicky\xspace}
\newcommand{\geu}{SN~iPTF16geu\xspace}

\newcommand{\gma}{\ensuremath{\gamma}\xspace}

\newcommand{\zd}{\ensuremath{z_d\xspace}}
\newcommand{\zs}{\ensuremath{z_s\xspace}}

\submitjournal{ApJ}

\shorttitle{SN Zwicky and SN iPTF16geu revisited}
\shortauthors{Baltasar, Ratier-Werbin, Huang et al.}

\begin{document}
\title{A Novel Lensed Point Source Modeling Pipeline using \gigal\\
with Application to SN~Zwicky and SN~iPTF16geu}

\correspondingauthor{Saul Baltasar, Nicolas Ratier-Werbin, Xiaosheng Huang}
\email{saul4@illinois.edu, nratier@ucm.es, xhuang22@usfca.edu}

\author[0009-0003-4697-7079]{Saul~Baltasar}
\altaffiliation{Equal coauthors.}
\affiliation{Physics Division, Lawrence Berkeley National Laboratory, 1 Cyclotron Road, Berkeley, CA 94720, USA}
\affiliation{Department of Physics, University of Illinois at Urbana-Champaign, Urbana, IL 61801, USA}

\author[0009-0009-8206-0325]{Nicolas~Ratier-Werbin}
\altaffiliation{Equal coauthors.}
\affiliation{Physics Division, Lawrence Berkeley National Laboratory, 1 Cyclotron Road, Berkeley, CA 94720, USA}
\affiliation{Department of Physics, 
Complutense University of Madrid, 28040 Madrid, Spain}

\author[0000-0001-8156-0330]{Xiaosheng~Huang}
\affiliation{Department of Physics \& Astronomy, University of San Francisco, San Francisco, CA 94117, USA}
\affiliation{Physics Division, Lawrence Berkeley National Laboratory, 1 Cyclotron Road, Berkeley, CA 94720, USA}

\author[0000-0003-1889-0227]{W.~Sheu}
\affiliation{Department of Physics \& Astronomy, University of California, Los Angeles, 430 Portola Plaza, Los Angeles, CA 90095, USA}
\affiliation{Physics Division, Lawrence Berkeley National Laboratory, 1 Cyclotron Road, Berkeley, CA 94720, USA}

\author[0000-0002-0385-0014]{C.~J.~Storfer}
\affiliation{Institute for Astronomy, University of Hawaii, Honolulu, HI 96822-1897, USA}
\affiliation{Physics Division, Lawrence Berkeley National Laboratory, 1 Cyclotron Road, Berkeley, CA 94720, USA}

\author[0000-0002-6876-8492]{Y.-M.~Hsu}
\affiliation{Department of Physics, National Taiwan University, No. 1, Section 4, Roosevelt Road, Taipei 106319, Taiwan}

\author{Sean~Xu}
\affiliation{Department of Physics, University of California, Berkeley, Berkeley, CA 94720, USA}
\affiliation{Physics Division, Lawrence Berkeley National Laboratory, 1 Cyclotron Road, Berkeley, CA 94720, USA}

\author[0000-0002-5042-5088]{David J. Schlegel}
\affiliation{Physics Division, Lawrence Berkeley National Laboratory, 1 Cyclotron Road, Berkeley, CA 94720, USA}

\begin{abstract}
\input{abstr}
\end{abstract}

\section{Introduction}\label{sec:intro}
\input{intro}

\FloatBarrier
\section{Formalism}\label{sec:formalism}
\input{formalism}
\xh{\section{Convergence Metrics}
\label{sec:converg-metrics}
\input{convergence}}

\section{Modeling results for simulated systems}\label{sec:sim}
\input{sim-systems}

\section{Application to two real systems}\label{sec:real-systems}
\input{real-systems.tex}

\section{Discussion}\label{sec:discussion}
\input{discussion}
\section{Conclusion}\label{sec:conclusion}
\input{conclusion}

\newpage
\section*{Acknowledgement}\label{sec:acknowledgement}

\software{
    TensorFlow \citep{TensorFlow},
    TensorFlow Probability \citep{dillon2017a}, 
    JAX \citep{bradbury2018a}, 
    Optax \citep{optax2020},
    lenstronomy \citep{birrer2018a},
    Matplotlib \citep{hunter2007a},
    photutils \citep{bradley2023a}
    seaborn \citep{waskom2021a},
    corner.py \citep{foreman2016a},
    NumPy \citep{harris2020a}
}

\bibliographystyle{aasjournal}
\bibliography{dustarchive}



\end{document}

%% file: abstr.tex
We introduce a novel modeling pipeline for strongly lensed point sources, using the \texttt{GIGA-Lens} framework, running on four A100 GPUs via the JAX platform. 
Using simulations, we demonstrate accurate and precise recovery of image positions, fluxes, and time delays, together with inference of complex lens mass distributions—including the mass density slope, $\gamma$—\emph{from images of lensed point sources alone.}
We further show that we can achieve statistical uncertainty of \xh{$\sim 3.6\%$ ($\sim 2.5\, \mathrm{km\, s^{-1}/Mpc}$)} on $H_0$ from a \emph{single} system, with full forward modeling, i.e., simultaneous inference of all lens model parameters together with \ho.
\xh{We apply our pipeline to two well-studied lensed SNe~Ia, Zwicky and iPTF16geu. 
For SN~iPTF16geu, unlike previous modeling efforts, we model \emph{only} the images of the lensed point source (the SN) and do not use the lensed images of the extended host-galaxy. Nevertheless,
we are able to infer all of the mass parameters modeled in earlier studies, and our best-fit values, including $\gamma$, are fully consistent with published results.
 In the case of SN~Zwicky, taking the same approach, however, we obtain an alternative best-fit model compared to published results, underscoring the importance of \emph{fully} exploring the model parameter space.}

%% file: intro.tex
Currently there is a significant discrepancy between late and early universe observations of the Hubble constant (\ho), the expansion rate of the universe at the present day \citep{planck2018a, freedman2021a, riess2022a}.  
These measurements span a range of approximately $5\sigma$.  
An important question to address therefore is whether this stems from systematic bias in one or more measurements.

Lensed time-varying events have the potential to resolving this tension. For such events (typically quasars or transients, such as supernovae) in a strongly lensed host galaxy, the time delays between the multiple images can be measured and used to constrain \ho, when accurately modeled \citep[e.g.][]{wong2019a, kelly2023a}.  
Lensed supernovae (SNe) are particularly valuable due to their well-defined light curves, and in the case of Type Ia (SN~Ia), their standardizable peak magnitudes. Lensed SNe can also be used to constrain the dark energy equation of state \citep[e.g.][]{pierel2021a, suyu2023a}.

However, lensed~SNe are very rare, with 10 confirmed cases reported.
Three live lensed SNe were found in galaxy-scale strong lensing systems \citep{goobar2017a, goobar2023a, pierel2023a, johansson2025, taubenberger2025}
and four in cluster lenses  \citep{kelly2015a, kelly2022a, frye2023a, suyu2025}.
Three were found retrospectively, one in galaxy scale lens \citep{quimby2014a} and two in cluster lenses \citep{rodney2021a, chen2022a}.
Finally, \citet{sheu2023a} presented seven retrospectively identified lensed SN candidates in galaxy-scale and small group lenses using data from the Dark Energy Spectroscopic Instrument (DESI) Legacy Imaging Surveys \citep{dey2019a}.

Compared to SNe, quasars have stochastic light curves.  However, lensed quasars are much more common.
 Using six quasars, \citet{wong2019a} was able to constrain \ho with a $2.4 \%$ precision with mass profile assumptions.  
 The systematic uncertainty however is estimated to be $\sim 10\%$ \citep[e.g.,][]{birrer2020a}, spanning the range of the ``Hubble tension''.
 The ability to model both kinds of time-varying events robustly can therefore maximize the utility of these rare events to constrain cosmological parameters. 
In addition to physical factors that can introduce bias to lens modeling (e.g., the mass sheet degeneracy), we show that another critical source of bias arises from the modeling process itself.
Our results highlight the advantage of full forward modeling in arriving at the true optimum.

In this work, we present a novel pipeline for modeling lensed point sources, using \gigal.
\gigal was presented in \citet{gu2022a}.
It so far has been was applied to two real strong lens system with ground-based imaging \citep{cikota2023a, urcelay2024a} and seven system with \hst data \citep{huang2025a, huang2025b}, with extended lensed sources in all three cases.
This paper is organized as follows.
\sbj{We present our formalism in \S\,\ref{sec:formalism}, and describe our convergence metrics in \S\,\ref{sec:converg-metrics}.
In \S\,\ref{sec:sim}, we show modeling results for five simulated archetypal lensing systems, along with results for an additional system used to infer \ho. We then apply our pipeline to two known lensed SNe (Zwicky and iPTF16geu) in \S\,\ref{sec:real-systems}, and we discuss our results in \S\,\ref{sec:discussion}. Finally, we provide a conclusion in \S\,\ref{sec:conclusion}.}

%% file: formalism.tex



We introduce our lens modeling framework for lensed point sources. This method implements three different terms to the loss function in a differentiable manner, which allows for modeling of a variety of systems that are at the frontier of cosmological and astrophysical research, including multiply-lensed SNe and quasars. 
SN host galaxy light 
is not used for the modeling process.
Our approach uses only point source data, that is, image positions, observed fluxes, and observed time delays.
To our knowledge, this is the first instance where all three types of information are simultaneously used as constraints in the modeling process. Furthermore, we show that point source data alone is sufficient to determine the mass slope $\gamma$ of the elliptical power law (EPL) model.

Our framework offers significant advantages 
due to differentiability, which substantially enhances computational speed, optimization, and sampling.
\sbj{It is built upon the \gigal framework \citep{gu2022a}, a lens modeling pipeline that leverages the power of GPUs and Automatic Differentiation (AD).
In this Bayesian framework, one first sets the prior distributions for the model parameters. 
The likelihood function is then defined; for extended sources---the original use case of \gigal---it is based on pixel-level differences between the model and the data. 
The posterior distribution is the product of the prior and the likelihood.
Optimization of the posterior comprises three stages: multi-start gradient descent to find the Maximum a Posteriori (MAP), Stochastic Variational Inference (SVI) to identify an optimal Gaussian surrogate for the posterior, and Hamiltonian Monte Carlo (HMC) for full sampling.
Through these three stages, \gigal has demonstrated significant advantages in terms of speed
and sampling quality.}
Implemented in both TensorFlow and JAX, these libraries enable seamless execution of our code on GPUs, significantly accelerating gravitational lens simulation and modeling,
particularly given that JAX supports distributed computing across multiple GPUs. As in \citet{gu2022a}, we use all four A100 GPUs on a GPU node at NERSC.

\subsection{Limitations for point sources}
Lensed galaxy-scale systems are typically modeled using parametric models for the lens mass and light as well as for the source light. Then, a pixel-by-pixel comparison is performed, minimizing the difference between a forward-modeled image and an observed image.
Lensed point sources, however, cannot be modeled using this approach, which only applies to cases where observed images exhibit a sufficiently continuous pixel value distribution across the cutout image used in lens modeling. 
In such cases, with an appropriate prior, the observed and modeled images are expected to overlap substantially and the gradients will be of reasonable sizes in the optimization process.
For lensed point sources (e.g., quasars and SNe), this approach is infeasible, as the forward-modeled image positions and the observed image positions generally will not overlap at the beginning of the optimization process. 
This results in gradients that are essentially zero in much of the parameter space and thus not informative for optimization.
Below we will present a different approach, using only the centroid
of each lensed image of a point source.
We will show that this leads to a new smooth loss function with informative gradients.

\subsection{The Loss Function}\label{sec:total-loss-function}

Our likelihood function has three terms: 
\begin{equation}\label{eqn:joint-loss}
    \log\mathcal{L}((\boldsymbol{x},\boldsymbol{y}) \mid \Theta) = \omega_D\log\mathcal{L}_D((\boldsymbol{x},\boldsymbol{y}) \mid \Theta) + \omega_F\log\mathcal{L}_F((\boldsymbol{x},\boldsymbol{y}) \mid \Theta) + \omega_{TD}\log\mathcal{L}_{TD}((\boldsymbol{x},\boldsymbol{y}) \mid \Theta),
\end{equation}
where $\omega_D$, $\omega_F$, $\omega_{TD}$, $\mathcal{L}_D$, $\mathcal{L}_F$, $\mathcal{L}_{TD}$ are the relative weights and the likelihoods corresponding to each term of the loss function: delensed distance in the source plane (\ref{sec:distance-loss}), flux (\ref{sec:flux-loss}) and time delay (\ref{sec:TD-loss}), while $(\boldsymbol{x},\boldsymbol{y})$\footnote{We use $\boldsymbol{x}$ and $\boldsymbol{y}$ in boldface to indicate a set of $\{x_i,y_i\}$, with the index $i$ covering all observed images.} denotes the angular position of lensed images, and $\Theta$ the model parameters. In this work, the lens mass distribution described by $\Theta$ is given by the EPL model. This model can be characterized by the surface mass density expressed in units of the critical density, commonly referred to as convergence,
\begin{equation}
\label{eqn:epl-model}
    \kappa(x_{lens}, y_{lens}) = \frac{3-\gamma}{2}\left(\frac{\theta_E}{\sqrt{q x_{lens}^2+y_{lens}^2 / q}}\right)^{\gamma-1},
\end{equation}
where $\theta_E$ is the Einstein radius, $\gamma$ is the mass slope, $(x_{lens}, y_{lens})$ are lens-centric coordinates, and $q$ is the axis-ratio.
In lens modeling, often $q$ and the position angle $\phi$ are reparametrized as eccentricities,
\begin{equation}\label{eqn:e1,e2-vs-q-phi}
    (\epsilon_1, \epsilon_2) = \frac{1-q}{1+q}\left(cos(2\phi), sin(2\phi)\right).
\end{equation}
The external shear is parametrized by $\gamma_{ext, 1}$ and $\gamma_{ext, 2}$.

\xh{During early development, we found that a loss function consisting only of the distance term can drive the sampler toward unphysical regions of parameter space that formally yield lower $\chi^2$ values but are inconsistent with the known ground truths of simulated systems. Moreover, this numerical bias is not fully mitigated by the addition of the flux loss term (\S\,\ref{sec:flux-loss}) if observational uncertainties are used directly to set the relative weights between the two loss terms. We find that the adoption of a weighting scheme such that the different posterior components contribute comparably to the posterior explored during the HMC sampling stage yields stable and accurate inference. Below we first introduce the three loss terms, and then discuss how the relative weights are determined starting from observational uncertainties (\S\,\ref{sec:weights}).}

\subsubsection{Source Plane Compactness Loss Function}
\label{sec:distance-loss}
The first term of our loss function utilizes the positions
 $(\boldsymbol{x},\boldsymbol{y})$
 as a constraint. 
This is similar to the approach in \citet{urcelay2024a} for modeling a group-scale strong lens.
We proceed as follows.
\sbj{We delens each observed lensed image $i$ to its source-plane position $\boldsymbol{\beta}((x_{i}, y_{i}), \Theta)$, which depends only on the mass parameters. The quantity $\overline{\boldsymbol{\beta}((\boldsymbol{x},\boldsymbol{y}), \Theta)}$ denotes the mean source position inferred from all images.}
Subsequently, we minimize the distance of all delensed positions to this mean source position, as shown in Equation~\ref{eqn:distance-loss}. \sbj{The compactness of the source-plane configuration should be maximized (the separations of the delensed positions are minimized), since all lensed images originate from a single source point.}
To select the brightest pixels we use point spread function (PSF) fitting to determine the image centroids and we feed them as input to the loss function,
\begin{equation}\label{eqn:distance-loss}
\log\mathcal{L}_{D}((\boldsymbol{x},\boldsymbol{y}) \mid \Theta) = \sum_{i} \left| \boldsymbol{\beta}((x_i, y_i), \Theta) - \overline{\boldsymbol{\beta}((\boldsymbol{x},\boldsymbol{y}), \Theta)}\right|^2.
\end{equation}

\noindent
Note that uncertainties are not incorporated at this stage; they are introduced later in \S\,\ref{sec:weights}.

Even though far-apart delensed positions in the source plane are penalized,
no particular average delensed position is preferred.
While the true source is located at a specific position,
the loss function itself does not impose any preference for a given location in the source plane.
As a result, the constraint provided by the position-based loss function alone is relatively weak, and minimizing the compactness of the delensed points can lead to non-physical optima.
For instance, in the EPL model, mass slopes $\gamma \sim 1$ or smaller represent a non-physical region of parameter space that is favored by this position-only loss function.
Hence, additional constraints need be imposed in order to recover \emph{physical} solutions.

We impose boundaries on the prior so as to restrict the potential optima to the physically expected region of parameter space.
Parameter values lying outside this region are assigned zero prior probability, corresponding to a log-prior of minus infinity.
However, because our inference relies on gradient-informed algorithms, hard prior boundaries can introduce numerical instabilities and computational pathologies near these boundaries.
To avoid these issues and maintain full differentiability of our pipeline, we opt for bijectors, which were also used in \citet{gu2022a}. 
\tblue{We map the unconstrained space $\mathbb{R}^d$, where $d$ is the number of parameters of the model, to the constrained space (also physical space) given by the support of the prior, $\text{supp}(p)$, using bijectors
\begin{equation}
    g:\mathbb{R}^d\rightarrow \text{supp}(p), \Tilde{\Theta}\mapsto \Theta.
\end{equation}
That is, $g(\Tilde{\Theta})=\Theta$. Hereafter we use the tilde notation to denote quantities in the unconstrained space, including probabilities.}
For the mass slope, $\gamma$, with support $(a,b)$, a convenient bijector is the sigmoid mapping
\begin{equation}
    g_\gamma : z \mapsto a + \dfrac{b-a}{1+e^{-z}}.
\end{equation}

These bijector functions are implemented in both TensorFlow and JAX, allowing us to sample over the unconstrained parameter space and map those values to the constrained space. Since the volume elements are different, the prior needs to include an extra factor involving the Jacobian of the bijector $g$, evaluated in $\Tilde{\Theta}$ (Equation~\ref{eqn:posterior}), while the likelihood remains unchanged.
It is with these restrictions, e.g., a lower bound of 1.5 for the mass slope\footnote{Mass slopes below this value are known but rare. We use the final sampling results to assess whether the lower bound should be relaxed.} motivated by the loss function's bias towards the non-physical value of 1, that the Bayesian approach gains relevance, yielding a final log-posterior given by
\begin{equation}\label{eqn:posterior}
   \log\Tilde{p}(\Tilde{\Theta} \mid (\boldsymbol{x},\boldsymbol{y})) = \log\mathcal{L}((\boldsymbol{x}, \boldsymbol{y}) \mid g(\Tilde{\Theta})) + \log p(g({\Tilde{\Theta}}) ) + \log\vert J(\Tilde{\Theta})\vert.
\end{equation}

\subsubsection{Flux Loss Function}
\label{sec:flux-loss}
\sbj{The second term of our loss function (Equation~\ref{eqn:joint-loss}) involves flux data. 
The observed flux of image~$i$, denoted by $F_i^\text{obs}$, can be written as
\begin{equation}
F_i^\text{obs} = f^{\text{true}} \, \mu_i,
\label{eq:flux_true}
\end{equation}
where $f^{\text{true}}$ is the intrinsic (unlensed) flux of the source, and
$\mu_i$ is the physical lensing magnification of image $i$, fully determined by the lens mass model.
As mentioned in \S\,\ref{sec:total-loss-function} the incorporation of flux information helps restrict our parameter space away from non-physical solutions.}

\sbj{The intrinsic flux for a SN~Ia can be determined within $10\%$ \citep[e.g., ][]{boone2021a}, due to the standardizability of their luminosity.
For these systems, observational data are often reported in terms of \emph{observed magnifications}
$\mu_i^{\text{obs}}$ \citep[e.g.][]{larison2024a}, defined by dividing the observed (magnified) flux $F_i^\text{obs}$ by the inferred intrinsic flux $f^{\text{obs}}$,
\begin{equation}
\mu_i^{\text{obs}} \equiv \frac{F_i^\text{obs}}{f^{\text{obs}}} =  A \, \mu_i.
\label{eq:mu_obs_def}
\end{equation}
where the second equality comes from substituting Equation~\eqref{eq:flux_true}, and we have introduced the dimensionless parameter
\begin{equation}
A \equiv \frac{f^{\text{true}}}{f^{\text{obs}}}.
\label{eq:amplitude_def}
\end{equation}}

In our inference method, $A$ is therefore treated as a free parameter, with a prior distribution of $\mathcal{N}(1,0.1)$ for SNe~Ia;\footnote{For non-Type Ia SNe and quasars, we may adopt two alternative strategies: either fitting the amplitude with a broader prior or using flux ratios as constraints (Ratier-Werbin et al. in prep.).} and it is fitted simultaneously with the lens mass parameters.
This formulation cleanly separates lensing geometry, encoded in the magnifications $\mu_i$, from the intrinsic flux of the source.



In order to model magnifications we use the following parametric function
\begin{equation}\label{eqn:magnification}
    \mu ((x, y), \Theta) = \frac{1}{\mathrm{det}(\mathcal{A}((x, y), \Theta))},
\end{equation}
where $\mathcal{A}(\boldsymbol{\theta}, \Theta) = \frac{\partial \boldsymbol{\beta}(\boldsymbol{\theta}, \Theta)}{\partial \boldsymbol{\theta}} = \left(\delta_{ij} - \frac{\partial^2 \psi(\boldsymbol{\theta}, \Theta)}{\partial \theta_i \partial \theta_j}\right)$
represents the Jacobian matrix, $\boldsymbol{\theta}=(x,y)$
denotes the angle between the arbitrarily chosen optical axis and positions in the image plane, $\psi$ is the effective lensing potential, and the delensed position, 
$\boldsymbol{\beta} = \boldsymbol{\theta} - \boldsymbol{\alpha}$, is determined by the deflection angle, $
\boldsymbol{\alpha} = \nabla \psi (\boldsymbol{\theta}, \Theta)$.
This function has a singularity along the critical curve and thus requires a special treatment in gradient-informed methods. For images close to the critical curve, minor adjustments in lens mass parameters can move the model so close to singularities as to causing computational errors. 
Manually restricting the parameter space near the critical curve to prevent singularities is infeasible without prior knowledge of the best-fit lens mass parameters.
Composing the magnification with a bounded and smooth function is ineffective due to gradient propagation through AD, where instabilities are transmitted via the chain rule. 

We therefore introduce a decomposition-based approach. We decompose the non-differentiable function, obtaining a smooth component with a direct correspondence to the observed data.
This method can be applied to any non-differentiable function of the form $h = g \circ f$, where $f$ is the differentiable component and $g$ is continuous at one point, corresponding to $f$ evaluated at the best-fit parameters. The function $g$ need not be 
continuous everywhere. Let $\tilde{h}$ be the observed value modeled by $h$. If $ g^{-1}$ (such that  $ g^{-1} \circ h = f$) can be explicitly determined, we can minimize $\|f - g^{-1}(\tilde{h}) \|$ using a gradient-informed method, as it involves only a differentiable function. The term $\|g\circ f - \tilde{h}\|$ can be made arbitrarily small by minimizing $\|f - g^{-1}(\tilde{h})\|$ due to the continuity of $g$ at the value of $f$ corresponding to the best-fit parameters.

In our case, we apply this method to the magnification function given in Equation~\ref{eqn:magnification}, considering that the denominator is differentiable and that its corresponding observed values can be obtained by inverting the observed magnifications.
Hence, our method fits the inverse observed magnifications, $1/\mu_i^\text{obs}$, using the model prediction $\det(\mathcal{A}((x_i, y_i), \Theta))/A$.
Since parity of individual images cannot be measured, we square the observed fluxes, yielding the following final form of the log-likelihood
\begin{equation}\label{eqn:flux-loss}
    \log\mathcal{L}_{F}((\boldsymbol{x},\boldsymbol{y}) \mid \Theta) = \sum_{i} \left| \left(\dfrac{\mathrm{det}(\mathcal{A}((x_i, y_i), \Theta))}{A}\right)^2 - \left(\dfrac{1}{\mu_i^{obs}}\right)^2 \right|^2.
\end{equation}
This loss term further restricts the parameter space, avoiding a non-physical region that source compactness minimization alone could not prohibit.

\sbj{One of the key innovations of our approach is the use of AD beyond just the optimization process. Specifically, we have incorporated AD\footnote{\ed{We employ forward-mode AD due to its slight advantage over reverse-mode when working with square matrices. Quoting from the \href{https://docs.jax.dev/en/latest/notebooks/autodiff_cookbook.html}{JAX documentation}, ``For matrices that are near-square, forward-mode automatic differentiation probably has an edge over reverse-mode."}} for the computation of magnifications (Equation~\ref{eqn:magnification}).
This enables the evaluation of nested derivatives entirely within AD---it is applied to the derivatives in the magnification calculation, and then applied again to propagate gradients of the objective function  during optimization.
Using AD for calculating magnifications, as opposed to finite differences, presents two main benefits. First, we observe a significant improvement in computational efficiency, with modeling speed improvements of approximately 25\%, 20\%, and 40\% in the MAP, SVI, and HMC steps, respectively. Second, while finite differences are sensitive to the choice of step size, AD ensures that derivatives are computed with machine-level precision. This is especially crucial in modeling lensed point sources, as these problems are highly non-linear---the solution approaches the critical curve singularity, where precision is paramount.}

\subsubsection[Time Delay Loss Function: \ho constraint]{Time Delay Loss Function: \ho constraint}
\label{sec:TD-loss}
Gravitationally lensed variable point sources provide yet another constraint on the lens model. 
Lensing causes unequal times of arrival for each image, known as the time delay, which vary depending on the structure of the mass distribution. Hence, we can use it as another constraint to our loss function.

In terms of the angular position of the source $\boldsymbol{\beta}_{src}(\Theta)$, and the effective lensing potential $\psi((\boldsymbol{x}, \boldsymbol{y}), \Theta)$, the Fermat potential is given by
\begin{equation}
    \Phi((\boldsymbol{x}, \boldsymbol{y}), \Theta) \equiv \Phi((\boldsymbol{x}, \boldsymbol{y}), \boldsymbol{\beta}_{src}(\Theta)) = \frac{1}{2} \left| (\boldsymbol{x}, \boldsymbol{y}) - \boldsymbol{\beta}_{src}(\Theta) \right|^2 - \psi((\boldsymbol{x}, \boldsymbol{y}), \Theta).
\end{equation}

Then, the time delay of image~$i$ can be written as
\begin{align}\label{eqn:time-delay}
    \Delta t_i((\boldsymbol{x}, \boldsymbol{y}), \Theta) &= \frac{D_{\Delta}}{c} \cdot \Phi((x_i, y_i), \Theta), \\
    \Delta t_{ij}((\boldsymbol{x}, \boldsymbol{y}), \Theta) &= \Delta t_i((\boldsymbol{x}, \boldsymbol{y}), \Theta) - \Delta t_j((\boldsymbol{x}, \boldsymbol{y}), \Theta),
\end{align}
where $\Delta t_{ij}$ denotes the time delay of image~$i$ relative to image~$j$, and $(x_i, y_i)$ the angular position of image~$i$. The time delay distance $D_{\Delta}$ is given by
\begin{equation} \label{eqn:time-delay-distance}
    D_{\Delta} = (1 + z_l) \frac{D_l D_s}{D_{ls}},
\end{equation}
with $z_l$ being the redshift of the lens and
$D_l$,
$D_s$,
$D_{ls}$ the angular diameter distance to the lens, to the source and from the lens to the source, respectively.

To avoid the factor of conversion in Equation~\ref{eqn:time-delay} in every iteration of the optimization process, we convert the observed time delays relative to the earliest image into \textit{observed} Fermat potential, using the inverse relation of Equation~\ref{eqn:time-delay}. This provides an equivalent loss function that involves fewer calculations. Rather than minimizing the difference between observed and predicted time delays, we minimize the difference of Fermat potentials
\begin{equation}\label{eqn:TD-loss}
    \log\mathcal{L}_{TD}((\boldsymbol{x},\boldsymbol{y}) \mid \Theta) = \sum_{i} \left| \left[\Phi((x_i, y_i), \Theta) - \Phi((x_0, y_0), \Theta)\right] - \Phi_{i,0}^{\text{obs}} \right|^2.
\end{equation}
Here, $\Phi_{i,0}^{\text{obs}}$ denotes the \textit{observed} Fermat potential of image~$i$ relative to the earliest observed image (indexed as $0$), $\Phi((x_i, y_i), \Theta)$ is the predicted Fermat potential at image~$i$, and we take the sum over all images.

While this modified loss function involving Fermat potential differences is computationally convenient, it assumes a known Hubble constant, \ho, to compute the time delay distance in Equation~(\ref{eqn:time-delay-distance}). We could also remove this assumption by minimizing for time delay differences instead. Hence, we would consider \ho as a fitting parameter and use the alternative loss function given by
\begin{equation}\label{eqn:TD-H0-loss}
    \log\mathcal{L}_{TD}((\boldsymbol{x},\boldsymbol{y}) \mid \Theta) = \sum_{i} \left| \Delta t_{i,0}((\boldsymbol{x}, \boldsymbol{y}), \Theta) - \Delta t_{i,0}^{\text{obs}} \right|^2.
\end{equation}

This time delay loss term provides an independent method to determine \ho; and indeed, in \S\,\ref{subsec:H0} we employ this form of the loss function to constrain \ho using a single system.

Generally, time delays scale with the Einstein radius.
Consequently, systems with very small Einstein radii can exhibit time delays shorter than one day, rendering any meaningful constraint on \ho effectively impossible. The two observed systems analyzed in this article fall into this regime: their measured time delays are not sufficiently long, and the associated uncertainties are comparable in magnitude to the measured values themselves. This limitation is consistent with their small Einstein radii, which we independently infer to be $0.176''$ and $0.289''$ in \S\,\ref{sec:real-systems}.
Meaningful \ho constraints are expected from systems with $\theta_E \gtrsim 1''$.

While the flux data helps restrict the parameter space to the physical domain more directly, we find that incorporating time delay measurements into the modeling process consistently improves three key areas. First, time delay data helps achieving sampling convergence, as indicated by a reduction in the Gelman-Rubin statistic, \rhat (see \S\,\ref{sec:converg-metrics}). Second, the uncertainties in the recovered parameters, particularly the mass slope $\gamma$, are reduced by up to 30\%. Third, while flux and positional data alone provide strong constraints on the mass slope, the inclusion of time delay information increases the accuracy of the $\gamma$ estimate. This results in a centroid that is closer to the ground truth.





\subsubsection{The Question of the Relative Weights}
\label{sec:weights}


Below we empirically determine the relative weights starting from observational uncertainties.
As stated at the end of Section \ref{sec:total-loss-function}, we adopt a weighting scheme to achieve accuracy and convergence.

We begin by providing a principled derivation of the relative weight $w_D$. The likelihood commonly used for source plane compactness maximization is
\begin{equation}\label{eqn:principled-loss}
    \sum_{i} \frac{\left| \boldsymbol{\beta}((x_i, y_i), \Theta) - \overline{\boldsymbol{\beta}((\boldsymbol{x},\boldsymbol{y}), \Theta)}\right|^2}{\sigma_i^2((x_i, y_i), \Theta)}.
\end{equation}
The error map 
comes from a first-order Taylor expansion of 
the deflection map, and is given by \citep[e.g.,][]{urcelay2024a}
\begin{equation}\label{eqn:error-map}
    \sigma_i^2((x_i, y_i), \Theta) = \mu ((x_i, y_i), \Theta)^{-2} \cdot \sigma_{\text{obs}, i}^2,
\end{equation}
where $\sigma_{\text{obs}, i}$ is the uncertainty in the position
$i$ and $\mu ((x_i, y_i), \Theta)$ is the magnification.
The model parameters that predict image positions lying close to the critical curve cause the magnification function to diverge and, as a consequence, the error map in Equation~\ref{eqn:error-map} approaches zero.
To regularize this behavior
we introduce an error floor, i.e., a minimum variance
\begin{equation}\label{eqn:error-floor}
    \sigma_{D,i}^2((x_i, y_i), \Theta) = \mu ((x_i, y_i), \Theta)^{-2} \cdot \sigma_{\text{obs}, i}^2 + \frac{1}{w_D}.
\end{equation}
This additional term $1/w_D$ prevents singularities and ensures numerical stability. Importantly, in the high-magnification limit, Equation~\ref{eqn:error-floor} naturally recovers the relative weight $w_D$ in Equation~\ref{eqn:joint-loss}. That is, we identify the meaning of $w_D$
as the error floor for the source plane compactness loss function. 

Although we derive the uncertainty for each individual image, in practice we adopt a single relative weight for each likelihood term. As a result, to interpret the relative weights in terms of observational uncertainties, we hereafter replace the image-level uncertainties by mean observed values obtained by averaging over all images.
In this way,  $w_D$ translates to a magnification function truncation,  
 $\mu_{\text{trunc}}$, via $\mu_{\text{trunc}}^{-1}\cdot \sigma_{\text{obs}} = (1/w_D)^{1/2}$, where $\sigma_{\text{obs}}$ is the average over all images.\footnote{When the first term in Equation~\ref{eqn:error-floor} vanishes, the variance becomes $1/w_D$. 
The truncated magnification that corresponds to this variance is $\mu_{trunc}$ 
such that $\mu_{trunc}^{-2}\sigma_{\text{obs}}^{2} = 1/w_D$.}

The relative weights $w_F, w_{TD}$ are determined directly from observational uncertainties.
Given that we fit for  $(1/\mu_i^{\text{obs}})^2$ 
rather than $(\mu_i^{\text{obs}})^2$,
the uncertainty in Equation~\ref{eqn:flux-loss} is given by
\begin{equation}\label{eqn:sigma-flux}
    \sigma_{1/\mu^2,i}^2 = \left(\frac{2 \sigma_{\mu,i}}{(\mu_i^{\text{obs}})^3}\right)^2.
\end{equation}
Here, $\sigma_{\mu,i}$ is the observed magnification uncertainty.\footnote{For clarity, let $\mu$ be the magnification with uncertainty $\sigma_{\mu}$. Performing error 
propagation for $y = 1/\mu^2$, $\sigma_y^2 = (dy/d\mu)^2 \sigma_\mu^2$. Since $dy/d\mu = -2/\mu^{3}$, then 
$\sigma_y^2 = (2\sigma_\mu/\mu^{3})^2$.}
Therefore, the relative weight $w_F$ relates to the uncertainty via $
w_F=1/2\sigma_{1/\mu^2}^2$, where $\sigma_{1/\mu^2}$ is the average over all images. Similarly, the relative weight $w_{TD}$ is given by $w_{TD} = 1/2\sigma_{TD}^2$, where $\sigma_{TD}$ denotes the average observed time delay uncertainty over all images.


Table
\ref{tab:relative-weights-simulation} demonstrates the consistency of our approach for a simulated system with uncertainties representative of real observations.
For simplicity, these experiments focus only on the position and flux contributions, as they provide the main constraints.
First, for a fixed value of $w_F$, we observe that increasing $w_D$---or equivalently, reducing the error floor---leads to consistent results. To interpret the $w_D$ values, we use the relation 
$\mu_{trunc}^{-1}\cdot \sigma_{\text{obs}} = (1/w_D)^{1/2}$, 
which links the error floor to an effective magnification truncation. Using $\sigma_{\text{obs}}$, we can map each $w_D$ to a corresponding $\mu_{trunc}$.
Second, using the relation 
$w_F=1/2\sigma_{1/\mu^2}^2$,
we can estimate $w_F$ from observational values. 
We find that accurately recovering observed positions, fluxes, and time delays and achieving convergence often requires increasing $w_F$ to values an order of magnitude higher than those estimated from observational uncertainties.
In practice, we begin with a relative weight estimated from observational uncertainties and gradually increase it until both accurate fits and convergence are achieved. We find a similar behavior for $w_{TD}$, which also requires a gradual increase.

The results in Table~\ref{tab:relative-weights-simulation} support the need to adjust the flux term weight.
As an illustrative case, we consider a system with ground truth $\gamma = 2.0$, and a prior given by a truncated Gaussian, with lower bound at $1.5$. 
For the low $w_F$ case, we see that the distance likelihood, based on image positions alone, favors solutions with low $\gamma$.
In fact, when using only the distance likelihood, we find that the posterior for $\gamma$ systematically converges to 1.0
in all experiments.
This represents the lowest physically plausible value, indicating a strong bias.
Therefore, although the ground truth value may still lie within the high-variance posterior, repeated modeling of different systems show that the mean is systematically underestimated.
In this case, we also observed inaccurate recovery of positions, fluxes, and time delays, as well as poor convergence.
In contrast, removing the distance likelihood and using only the flux likelihood prevents proper convergence.
We also test the opposite scenario, verifying that large $w_F$ values do not lead to overestimation bias for systems with low $\gamma$. Specifically, we model a system with ground truth $\gamma = 1.3$ and set $w_F = 5 \cdot 10^8$. The inferred posterior remains centered on the true value, with best-fit $\gamma = 1.299^{+0.025}_{-0.020}$, showing no evidence of overestimation bias.

\begin{deluxetable}{cccccccccccc}[h]
\tabletypesize{\scriptsize}
\label{tab:relative-weights-simulation}
\tablecaption{Best-fit parameters obtained with different relative weights for a simulated system.
We test different values of the distance likelihood weight $w_D$, interpreted via the effective magnification truncation $\mu_{trunc}$, while fixing the flux weight $w_F$ in the first two rows.
For low $w_F$, as shown in the last row, the recovered $\gamma$ is underestimated, confirming the need for proper weighting to avoid bias. Each configuration is evaluated based on whether the recovered fluxes and positions are consistent with observations and convergence
is achieved. All models share the same ground truth, allowing direct comparison. 
}
\setlength{\tabcolsep}{1.0pt}
\tablehead{
    \colhead{$w_D$} & 
    \colhead{$w_F$} &
    \colhead{$\mu_{trunc}$} &
    \colhead{$\theta_E$} &
    \colhead{$\gamma$} &
    \colhead{$\epsilon_1$} &
    \colhead{$\epsilon_2$} &
    \colhead{$x$} &
    \colhead{$y$} &
    \colhead{$\gamma_{\mathrm{ext},1}$} &
    \colhead{$\gamma_{\mathrm{ext},2}$} &
    \colhead{$A$}
}
\startdata
$10^7$ & $5\cdot 10^5$ & $9.5$& $0.16722_{-0.00030}^{+0.00049}$& $1.95_{-0.12}^{+0.13}$& $0.027_{-0.040}^{+0.039}$& $0.082_{-0.034}^{+0.030}$  & $0.00495_{-0.00062}^{+0.00061}$ & $0.01120_{-0.00070}^{+0.00090}$& $0.065_{-0.023}^{+0.020}$& $0.021_{-0.022}^{+0.025}$ & $0.984_{-0.098}^{+0.098}$ \\[4pt]
$5\cdot10^7$ &$5\cdot 10^5$ & $21$ & $0.16712_{-0.00020}^{+0.00047}$ & $1.94_{-0.11}^{+0.12}$  & $0.031_{-0.039}^{+0.039}$  & $0.088_{-0.018}^{+0.021}$  & $0.00508_{-0.00029}^{+0.00029}$ & $0.01124_{-0.00058}^{+0.00077}$ & $0.068_{-0.022}^{+0.019}$  &$0.019_{-0.011}^{+0.015}$ & $0.985_{-0.097}^{+0.098}$\\[4pt] 
$5 \cdot 10^7$ & Low $w_F\!:\!10^4$ & $21$ & $0.16724_{-0.00042}^{+0.00076}$& $1.78_{-0.19}^{+0.24}$ &
$0.036_{-0.051}^{+0.055}$&
$0.046_{-0.093}^{+0.060}$&
$0.00469_{-0.00137}^{+0.00088}$&
$0.0106_{-0.0015}^{+0.0014}$& $0.057_{-0.031}^{+0.032}$& 
$0.003_{-0.053}^{+0.036}$ & $1.0_{-0.1}^{+0.1}$\\ \hline
& Ground truth & & $0.167$ & $2.000$ & $0.049$  & $0.083$ & $0.005$ & $0.011$ & $0.078$ & $0.015$ & $1.000$\\ 
\enddata
\end{deluxetable}

\begin{figure}[h]
    \centering
    \includegraphics[keepaspectratio=true,scale=0.28]{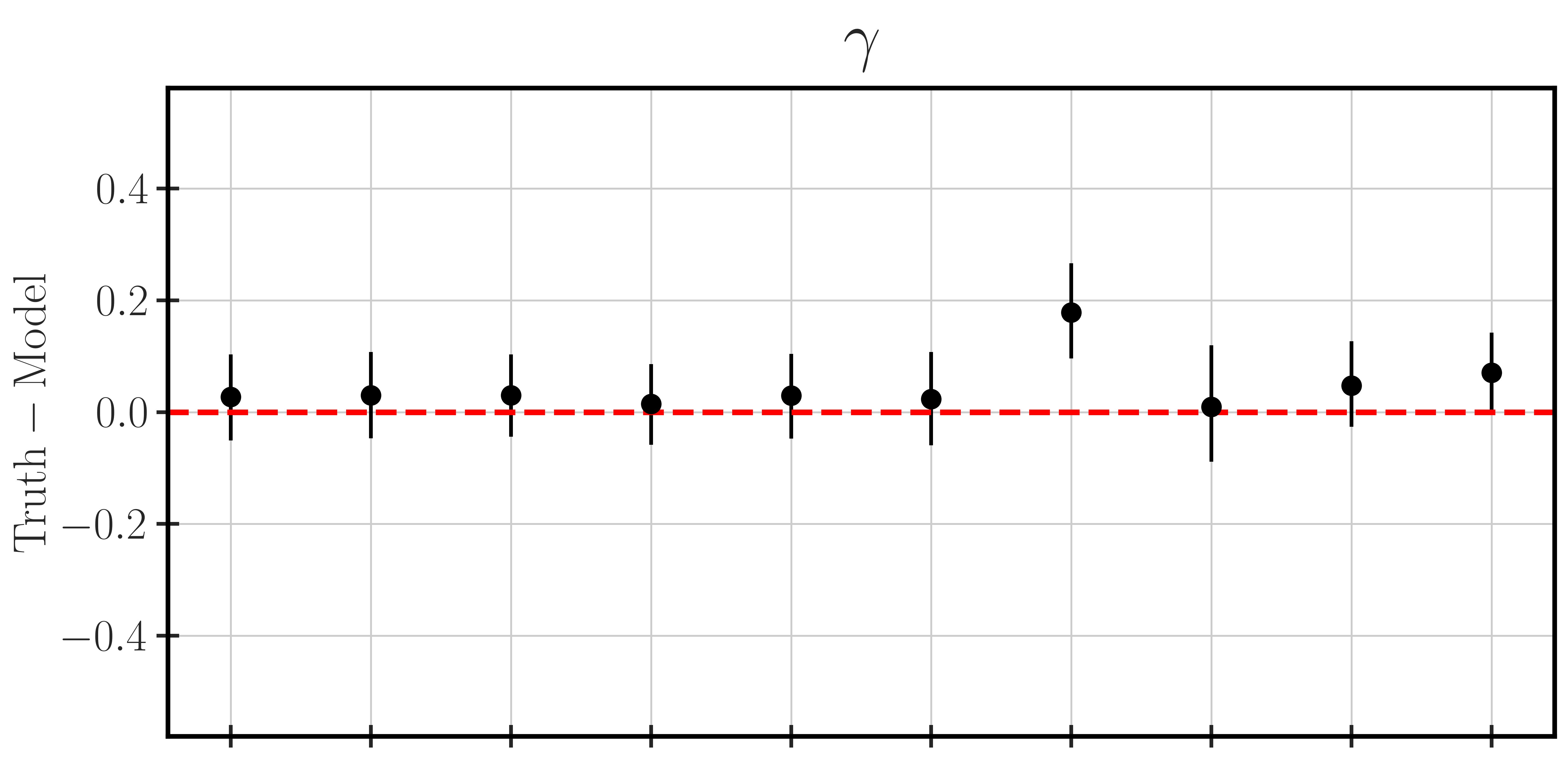}
    \caption{Difference between the ground truth and the best-fit $\gamma$ for 10 simulations. We use ground truth  $\gamma$ values sampled from $\mathcal{N}(2.0, 0.1)$. Error bars represent the 68\% posterior density intervals obtained via HMC sampling. Averaging over all systems, we find a mean error of --$0.045$ and a normalized mean error of --$0.56\pm0.57$ (i.e., $\sim 0.5\, \sigma$), indicating consistency with zero bias.}
    \label{fig:residual-gamma}
\end{figure}

Finally, to test the robustness of our method, we model ten systems with ground truth values of $\gamma$ sampled from $\mathcal{N}(2.0, 0.1)$. We use the same relative weights for all systems. While a separate hyperparameter (i.e., the weights) search for each system would produce even more accurate parameter estimates, we show that a single set of weights yield unbiased results. 
If the distribution used to generate ground truth values were broader, the fixed weights may no longer be valid. In that case, we observe two outcomes: either the ground truth is still recovered, or the model fails to converge. In practice, we perform a parameter search for the weights. 
Figure~\ref{fig:residual-gamma} shows the results across all systems. 
For each system, we compute the error as the difference between the posterior median and the truth. The normalized error is given by this difference divided by the posterior standard deviation.
Averaging over the ten systems, we find a mean error of $-0.045$ and a normalized mean error of $-0.56 \pm 0.57$, i.e., approximately half of a standard deviation. These results indicate consistency with zero bias.

\newpage
\sbj{Eliminating mathematical singularities in the magnification function, together with the selection of appropriate relative weights, enables both accurate fitting of positions, fluxes, and time delays and achieving convergence.}
\sbj{To avoid manually tuning the weights, a possible future extension of this work would be to treat them as fitting parameters, with priors centered on empirically motivated values.} 

%% file: convergence.tex
Before demonstrating the performance of our pipeline on simulated systems in the next section, we introduce the convergence metrics used throughout this work.

We employ two widely used diagnostics: the potential scale reduction factor, $\hat{R}$ \citep{gelman1992a}, and the effective sample size (ESS). The ESS provides an estimate of the number of effectively independent samples obtained from Hamiltonian Monte Carlo (HMC), with higher values indicating better sampling performance.

\citet{gelman1992a} suggested a convergence threshold of $\rhat < 1.1$ \citep[see also][]{gelman2014a}.\footnote{\ed{A discrepancy exists between the current definition of $\hat{R}$ \citep{gelman2014a} and its implementation in TensorFlow Probability \citep{brooks1998a}: the former is effectively the square root of the latter; see \citet{huang2025b} for details.}} 
We meet this criterion for all systems modeled in this work, including both simulated and real lenses.
More recently, \citet{vehtari2021a} recommended a more stringent criterion of $\rhat < 1.01$. They further emphasized the importance of accounting for cross-chain variance: sampling multiple chains is not equivalent to simply summing the ESS across chains, and neglecting this effect can lead to overestimation of the effective sample size.

This accounting is implemented in TensorFlow Probability by appropriately setting the \texttt{cross\_chain\_dims} argument (rather than using the default value of \texttt{None}). With this definition of ESS, \citet{vehtari2021a} recommend achieving $\mathrm{ESS} \gtrsim 100$ per chain for the corresponding $\rhat$ values to be considered reliable. In this work, we aim to meet these more stringent criteria.

All reported ESS values are calculated using this definition, and in all cases we satisfy the recommended threshold. For $\rhat$, we achieve $\rhat < 1.01$ for all simulated systems except the double,
and for both observed lenses; these cases are discussed further in \S\,\ref{sec:discussion}.


%% file: sim-systems.tex
To demonstrate the performance of our pipeline, we conduct modeling results on five simulated archetypal configurations (\S\,\ref{subsec:archetypal}), as well as an inference of \ho using a simulated system (\S\,\ref{subsec:H0}).

\subsection{Modeling Archetypal Systems}\label{subsec:archetypal}
\begin{figure}[H]
    \centering
    \includegraphics[keepaspectratio=true,scale=0.30]{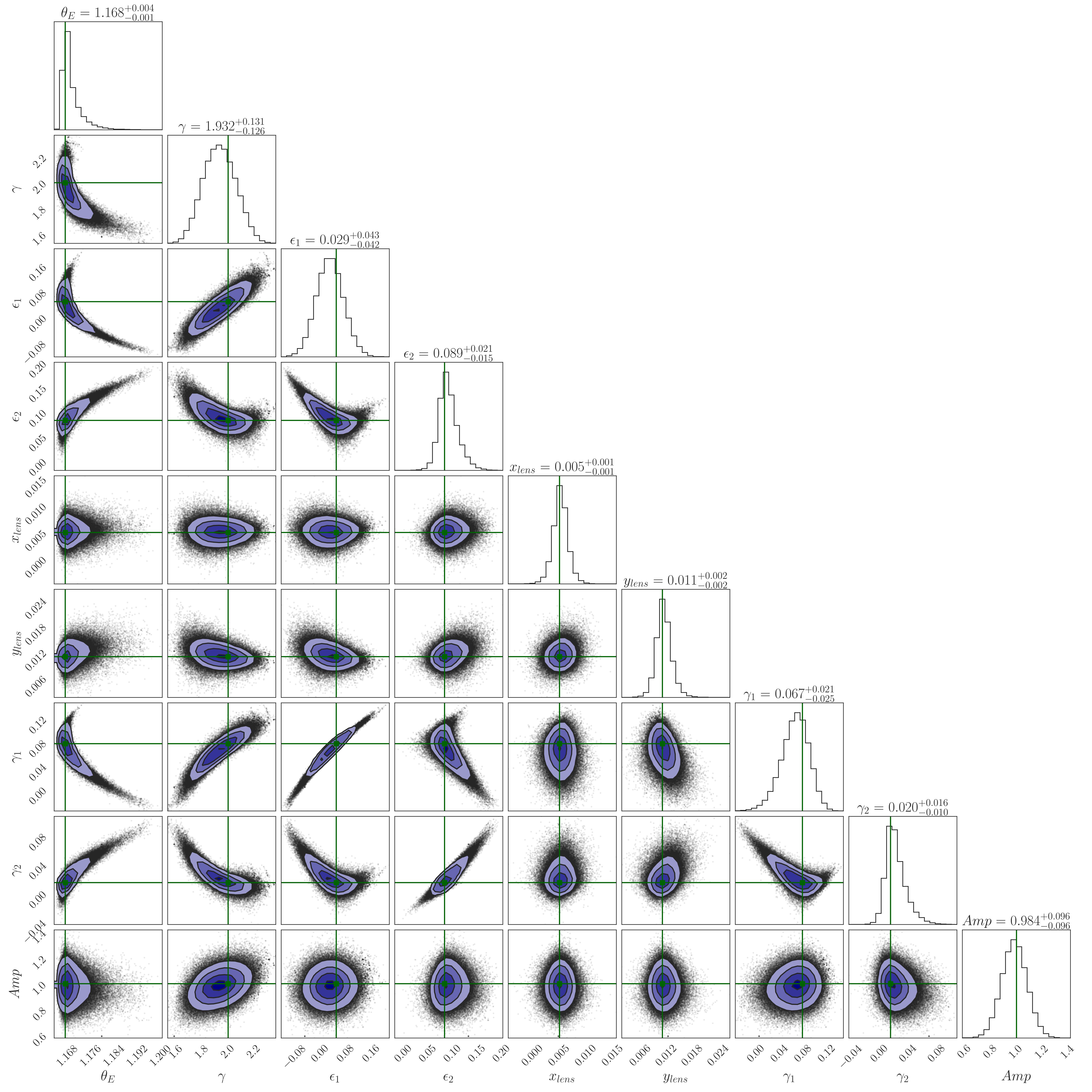}
      \begin{tikzpicture}[remember picture,overlay]
      \node at (-3.0cm,14.0cm) 
      {\includegraphics[scale = 0.40]{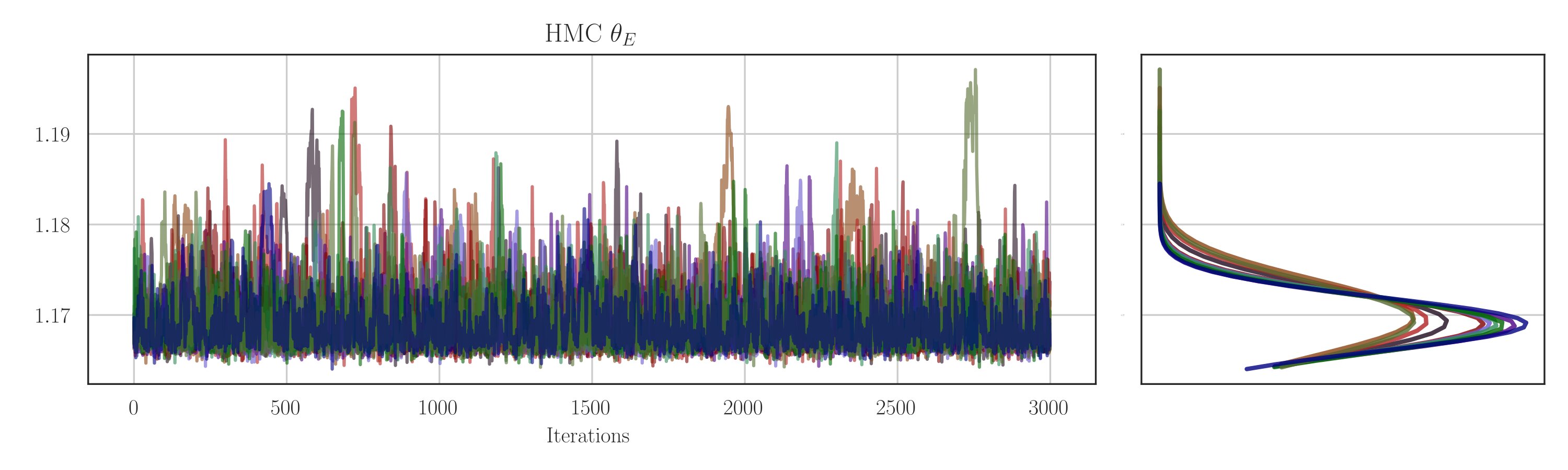}};
    \end{tikzpicture}
    \begin{tikzpicture}[remember picture,overlay]
      \node at (-3.1cm,11.0cm) 
      {\includegraphics[scale = 0.40]{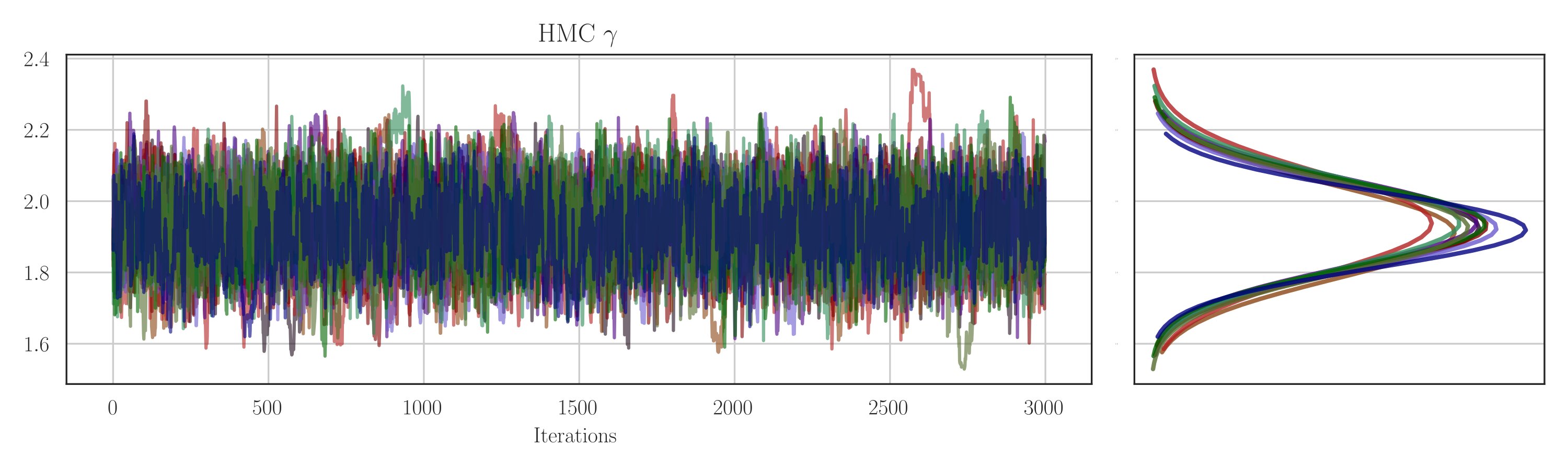}};
    \end{tikzpicture}
    \\\includegraphics[keepaspectratio=true,scale=0.40]{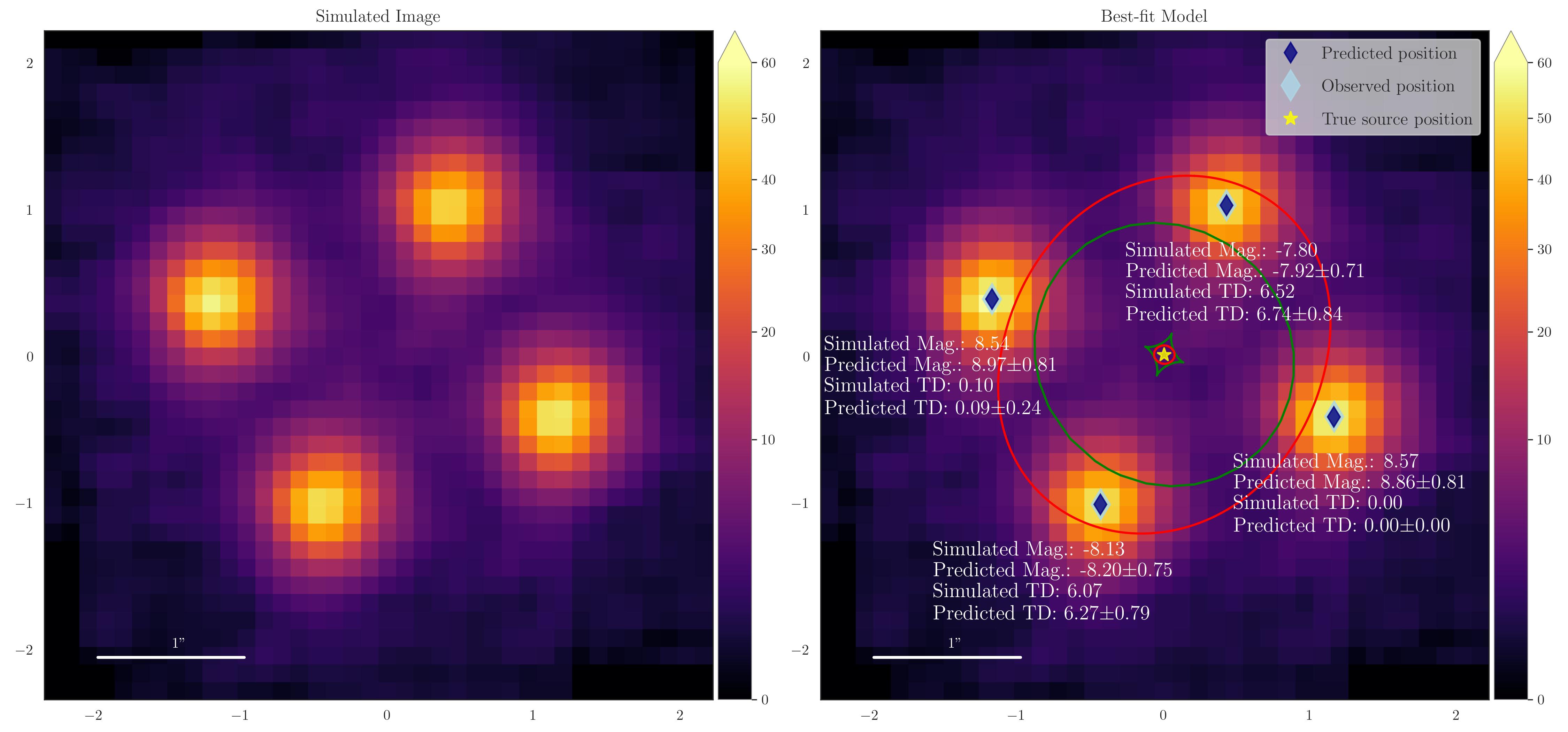}
    
    \caption{Modeling results for the cross configuration. The sampling converges with $\rhat_{max} = 1.004$ and \mbox{ESS $= 6366-134623$}, within 23 min.\ and 19 sec.\ of modeling time. Notice the banana-shaped marginals for some parameter combinations, indicating non-Gaussianity.}
    \label{fig:cross}
\end{figure}

\begin{figure}[H]
    \centering
    \includegraphics[keepaspectratio=true,scale=0.30]{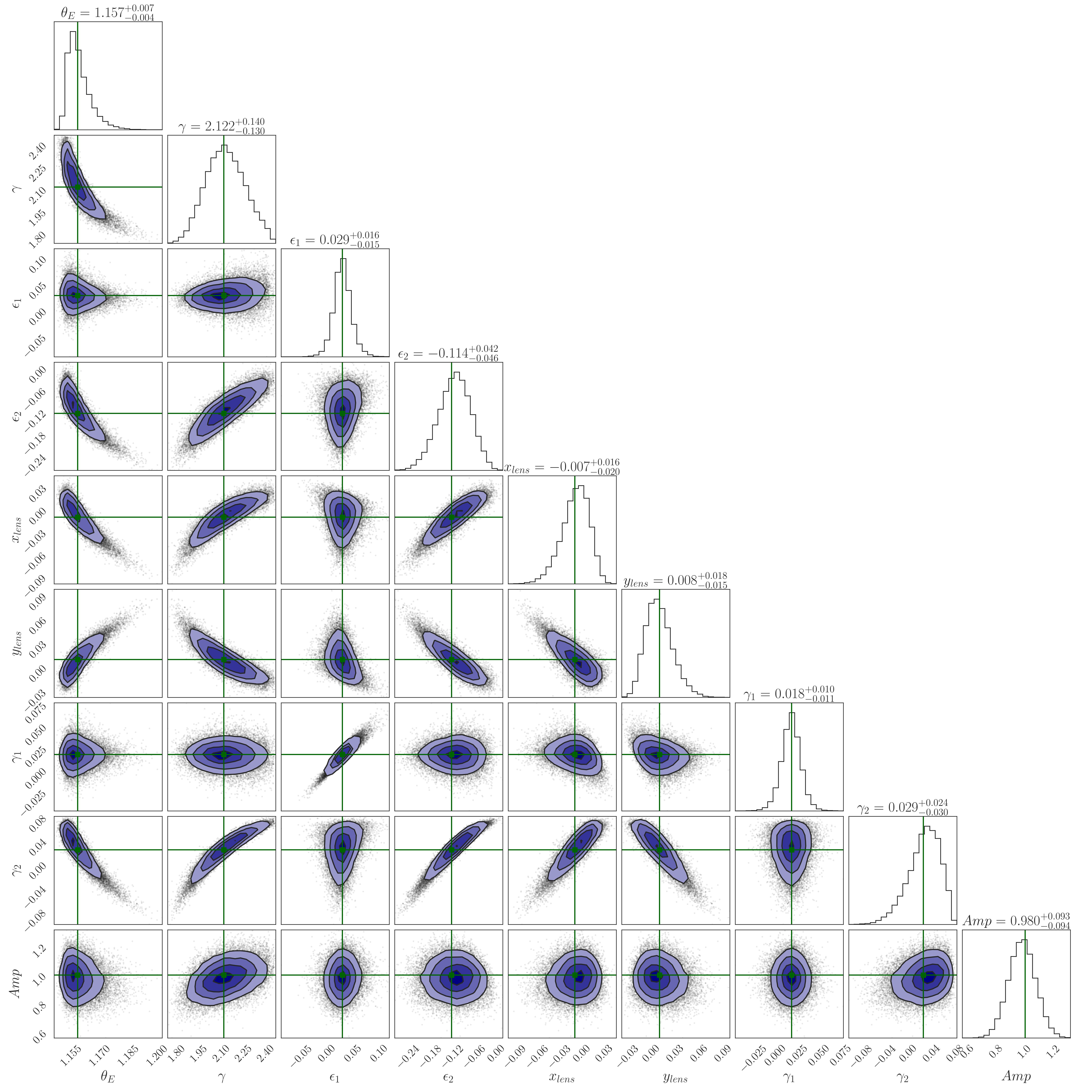}
      \begin{tikzpicture}[remember picture,overlay]
      \node at (-3.0cm,14.0cm) 
      {\includegraphics[scale = 0.40]{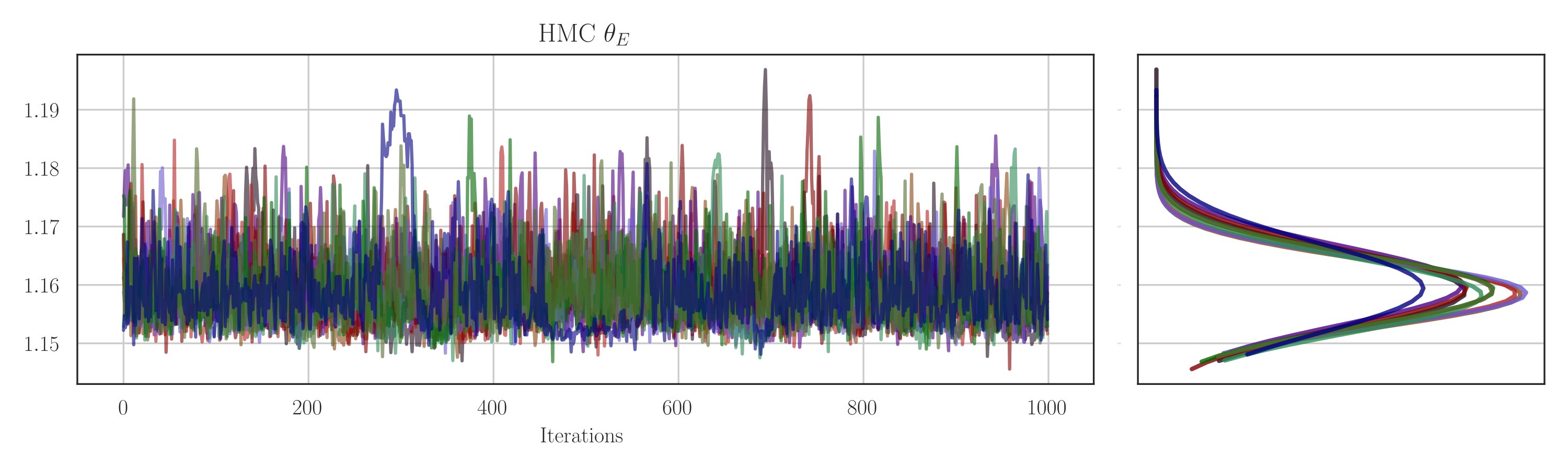}};
    \end{tikzpicture}
    \begin{tikzpicture}[remember picture,overlay]
      \node at (-3.1cm,11.0cm) 
      {\includegraphics[scale = 0.40]{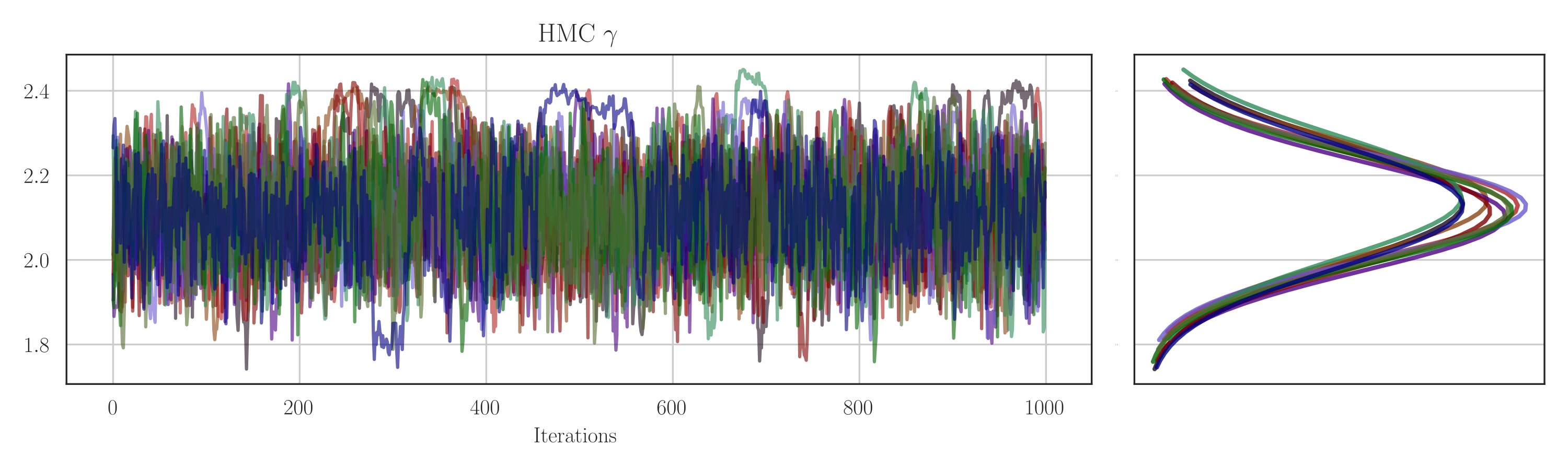}};
    \end{tikzpicture}
    \\\includegraphics[keepaspectratio=true,scale=0.40]{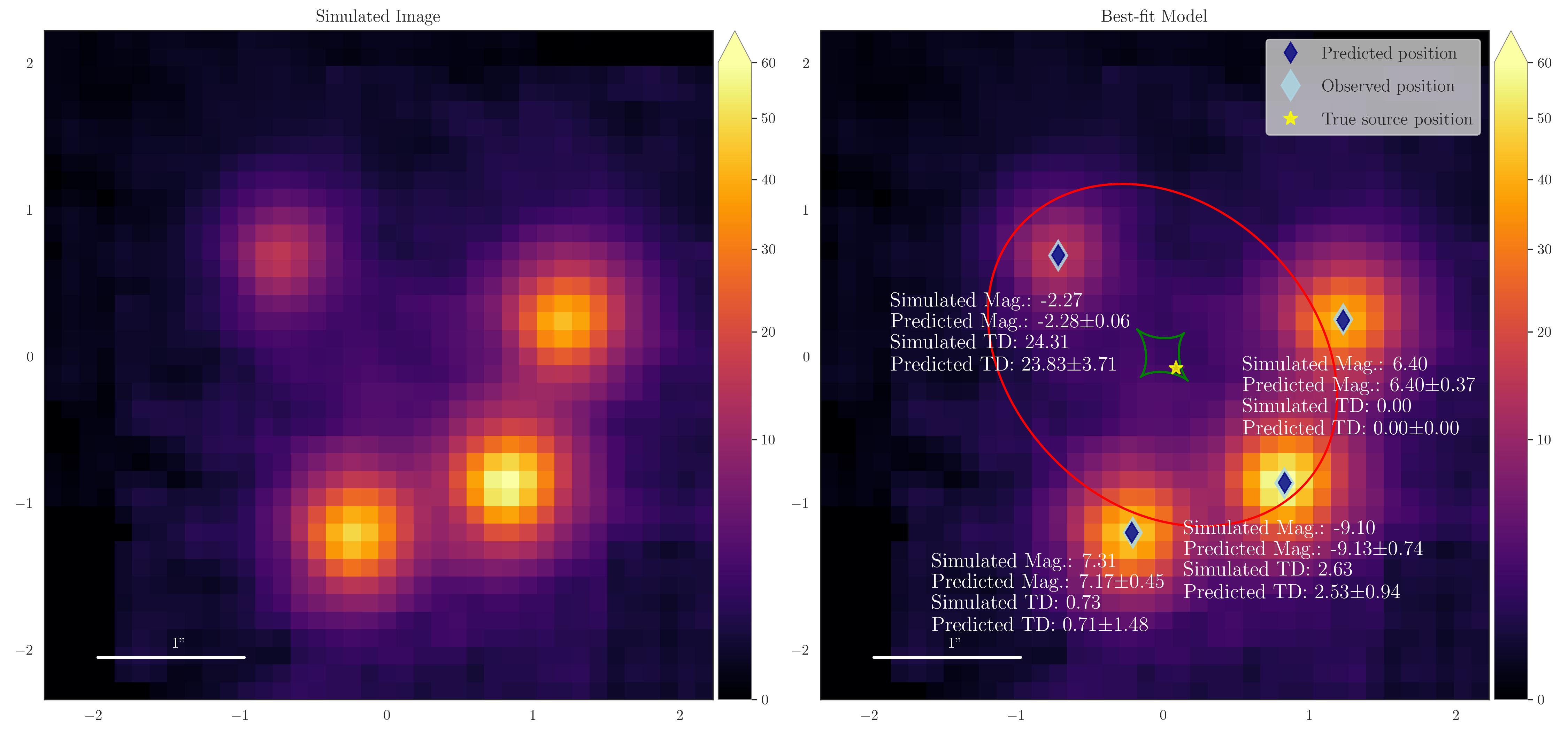}
    
    \caption{Modeling results for the long-cusp configuration. The sampling achieves $\rhat_{max} =  1.005$ and \mbox{ESS $= 1834-19525$}, with a total modeling time of 5 min.\ and 25 sec.\
    }
    \label{fig:long-cusp}
\end{figure}

\begin{figure}[H]
    \centering
    \includegraphics[keepaspectratio=true,scale=0.30]{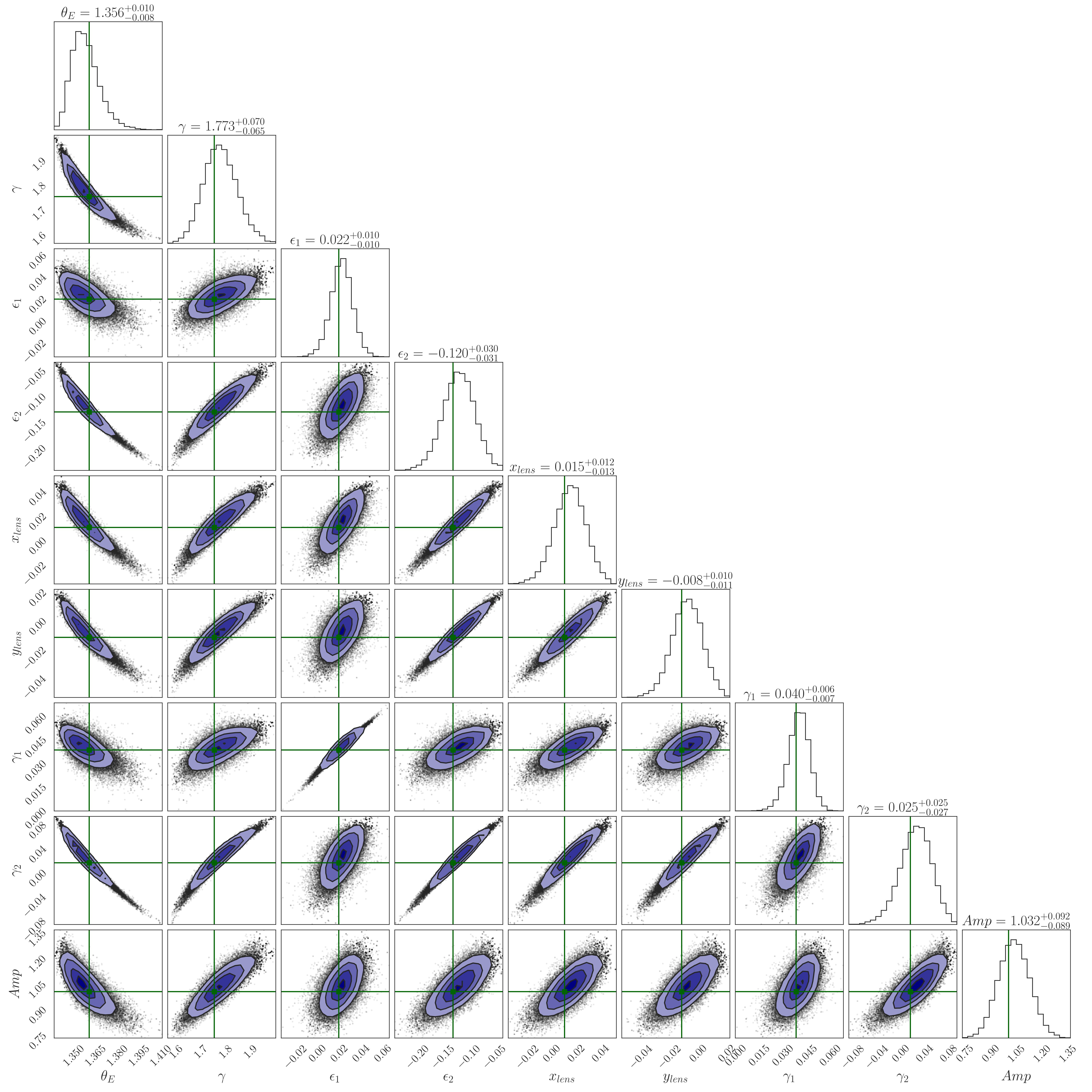}
      \begin{tikzpicture}[remember picture,overlay]
      \node at (-3.0cm,14.0cm) 
      {\includegraphics[scale = 0.40]{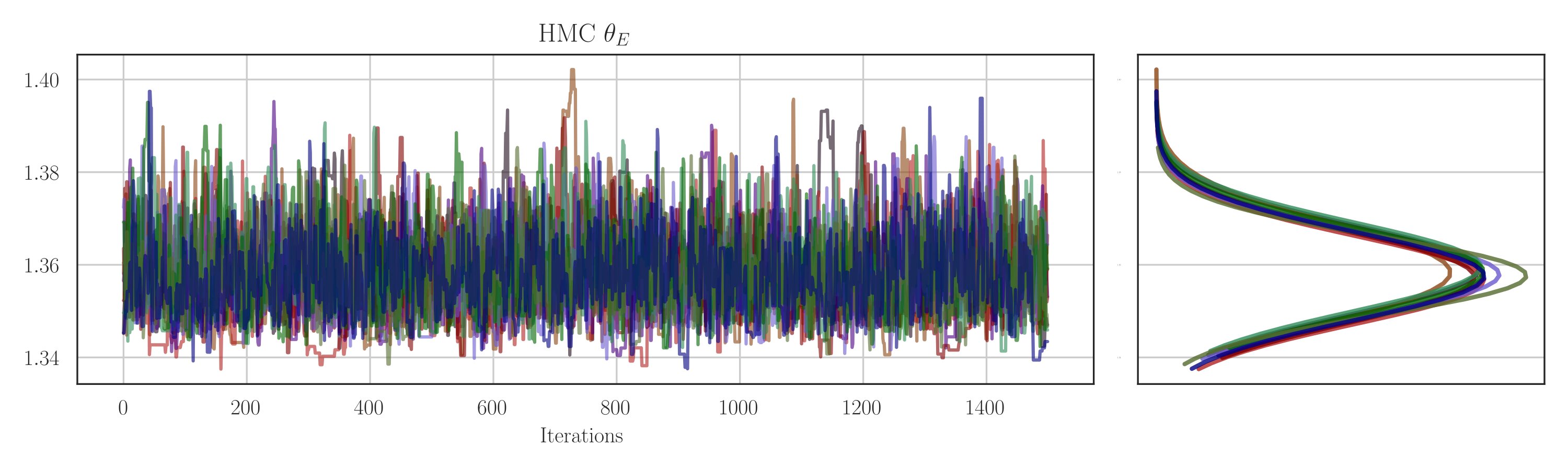}};
    \end{tikzpicture}
    \begin{tikzpicture}[remember picture,overlay]
      \node at (-3.1cm,11.0cm) 
      {\includegraphics[scale = 0.40]{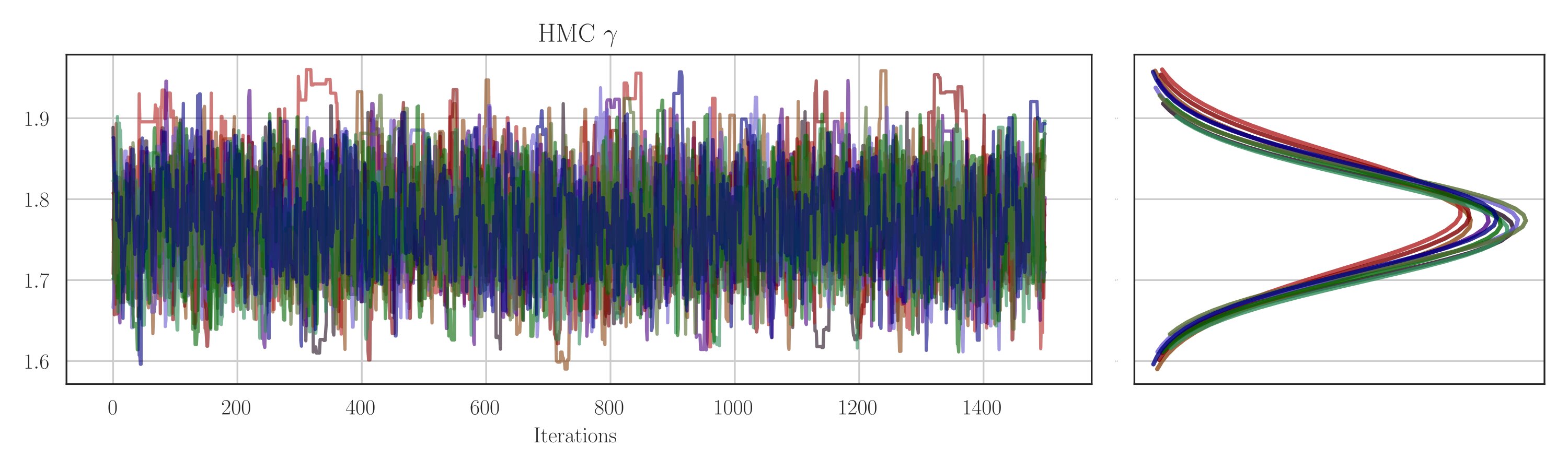}};
    \end{tikzpicture}
    \\\includegraphics[keepaspectratio=true,scale=0.40]{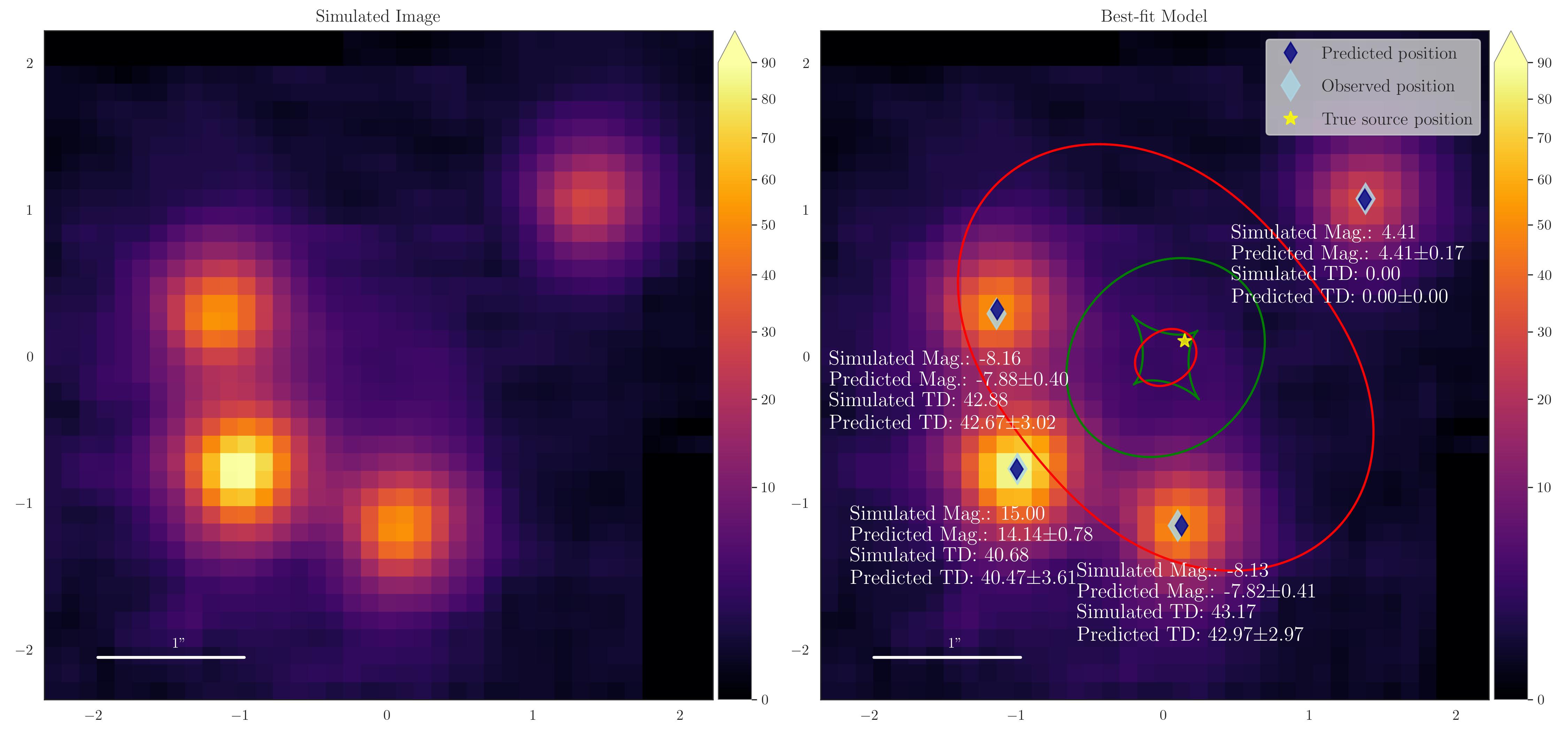}
    
    \caption{Modeling results for the short-cusp configuration. The sampling achieves $\rhat_{max} =  1.004$ and \mbox{ESS $= 4691-7132$}, within 2 min.\ and 1 sec.\ of modeling time. Notice the presence of an inner critical curve and caustic due to $\gamma < 2$.}
    \label{fig:short-cusp}
\end{figure}

\begin{figure}[H]
    \centering
    \includegraphics[keepaspectratio=true,scale=0.30]{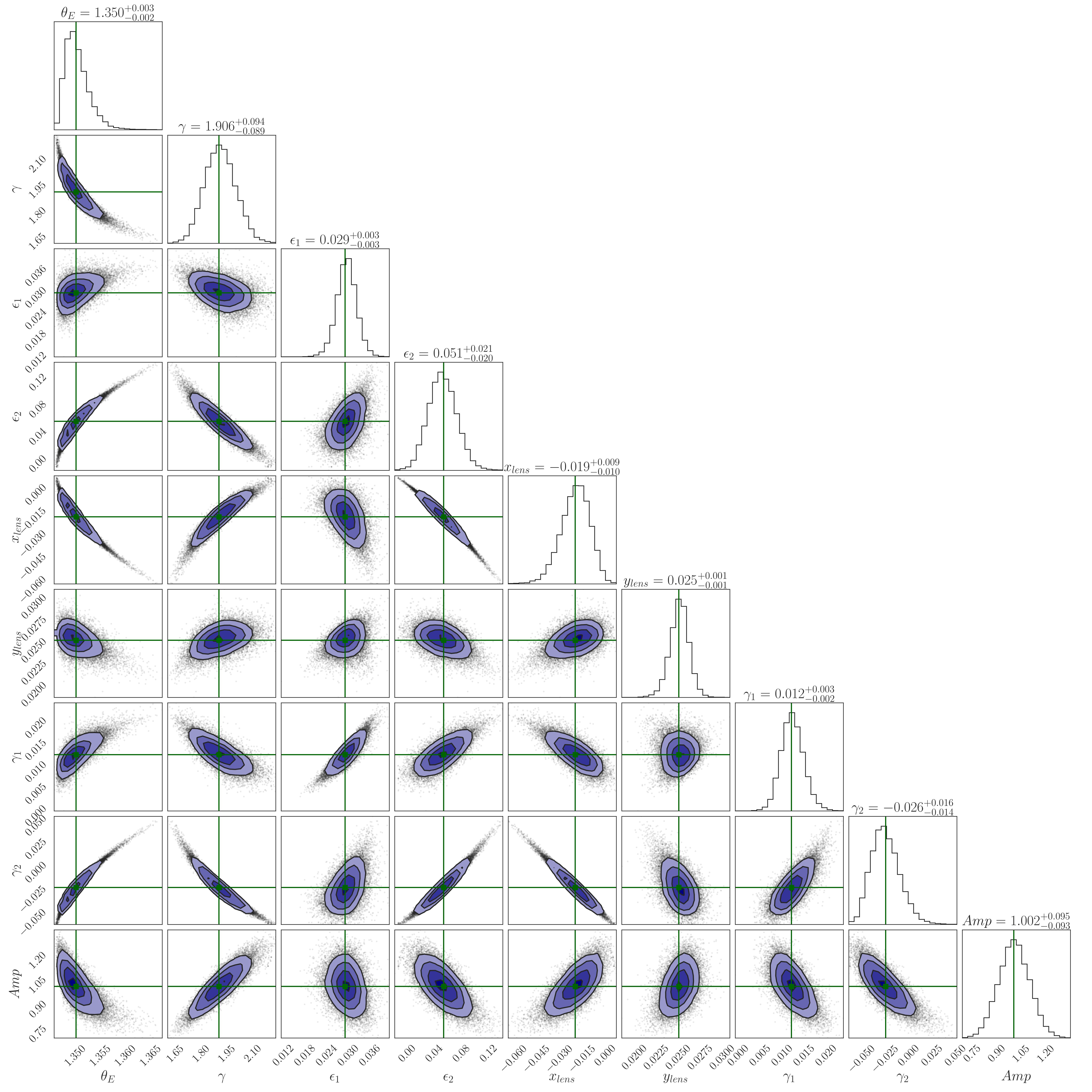}
      \begin{tikzpicture}[remember picture,overlay]
      \node at (-3.0cm,14.0cm) 
      {\includegraphics[scale = 0.40]{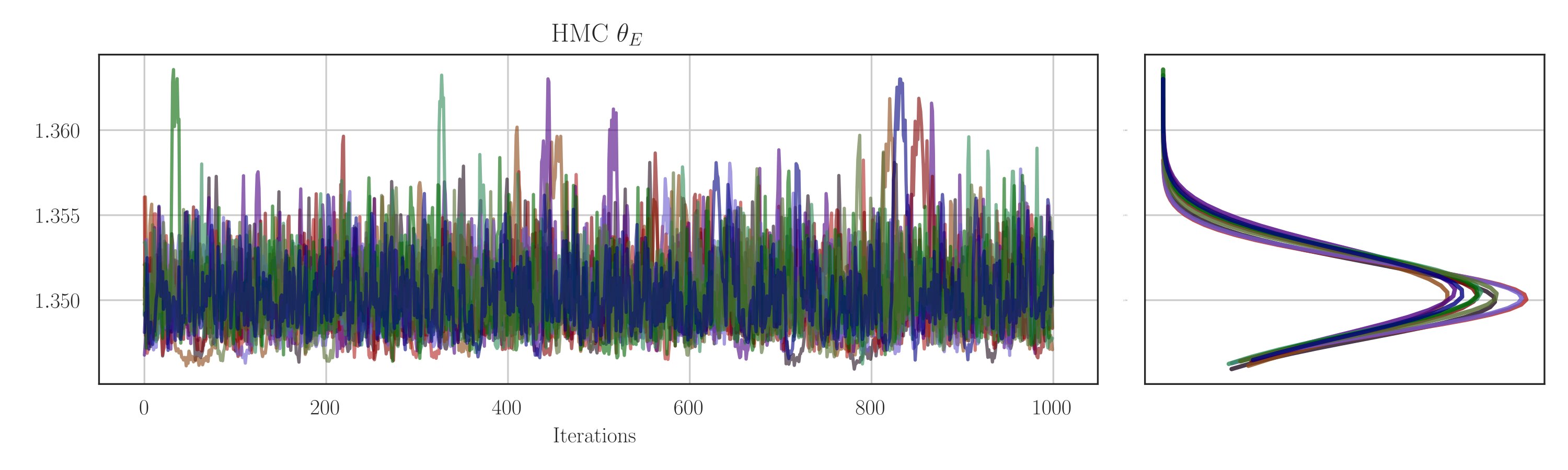}};
    \end{tikzpicture}
    \begin{tikzpicture}[remember picture,overlay]
      \node at (-3.1cm,11.0cm) 
      {\includegraphics[scale = 0.40]{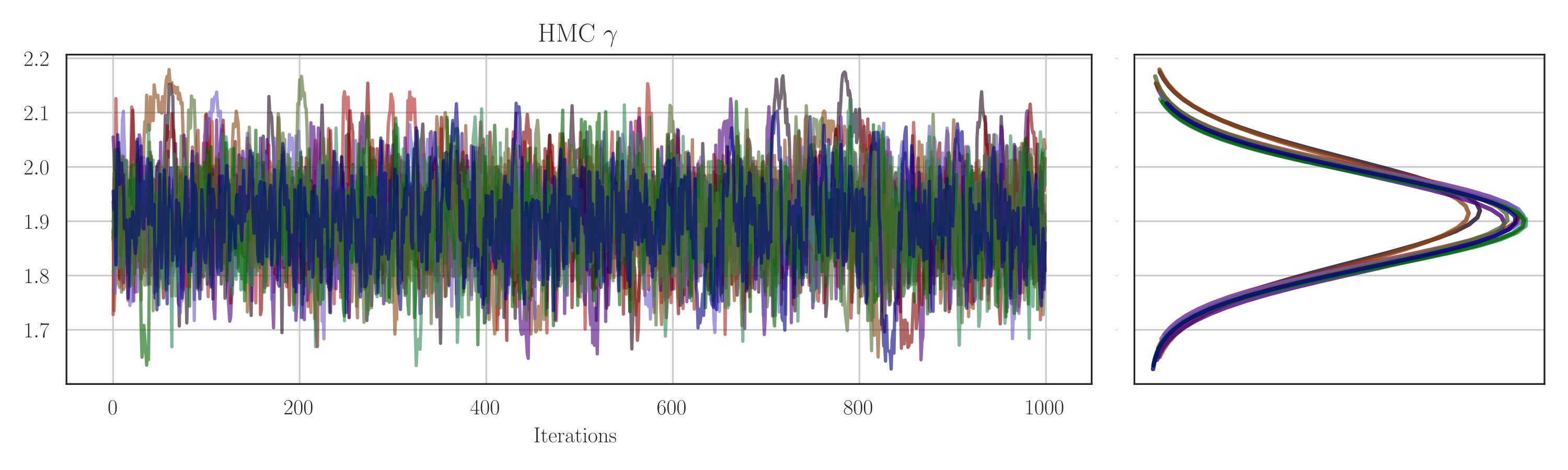}};
    \end{tikzpicture}
    \\\includegraphics[keepaspectratio=true,scale=0.40]{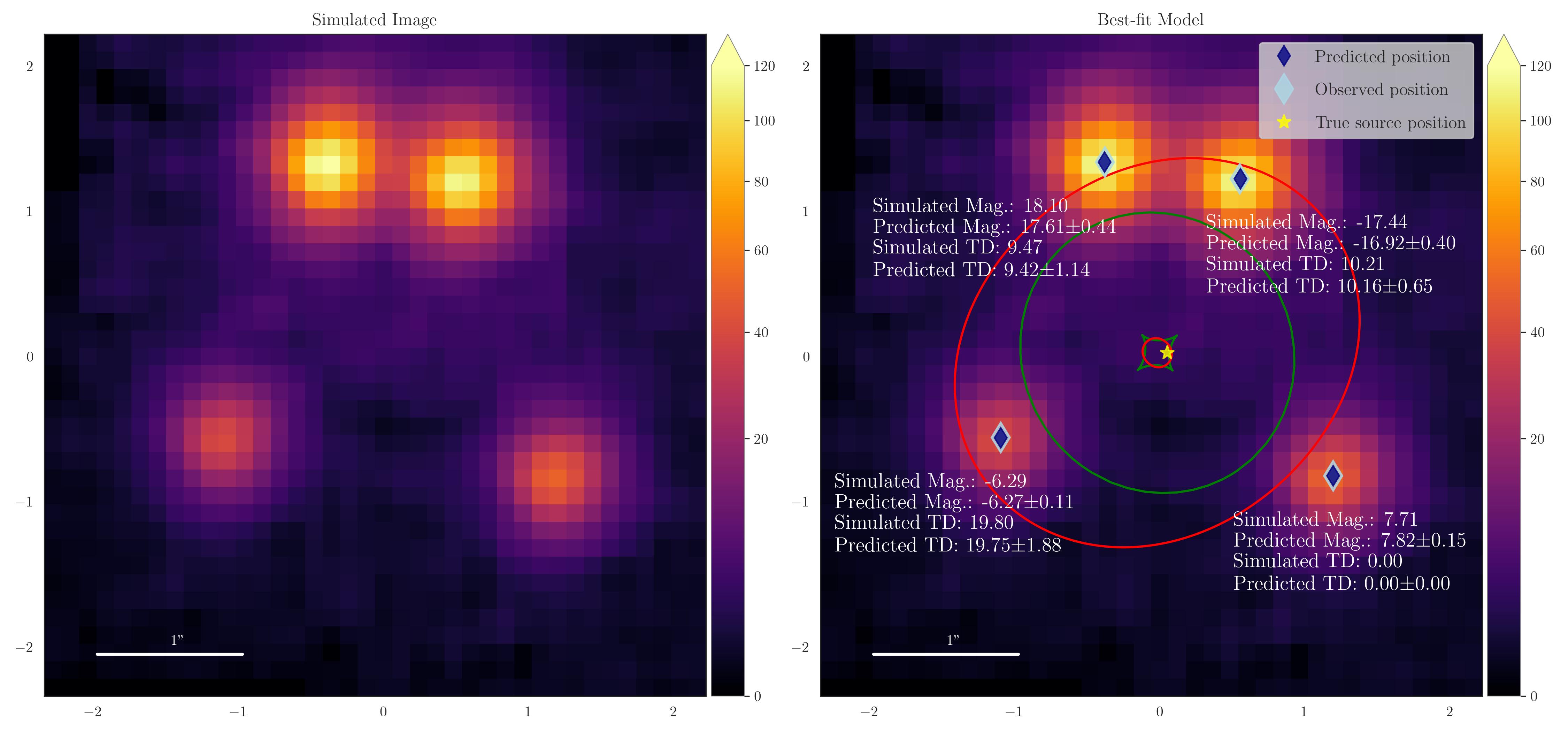}
    
    \caption{Modeling results for the fold configuration. The sampling results yield $\rhat_{max} =  1.002$ and \mbox{ESS $= 2546-11008$}, for a modeling time of 22 min.\ and 14 sec.\ } 
    \label{fig:fold}
\end{figure}

\begin{figure}[H]
    \centering
    \includegraphics[keepaspectratio=true,scale=0.30]{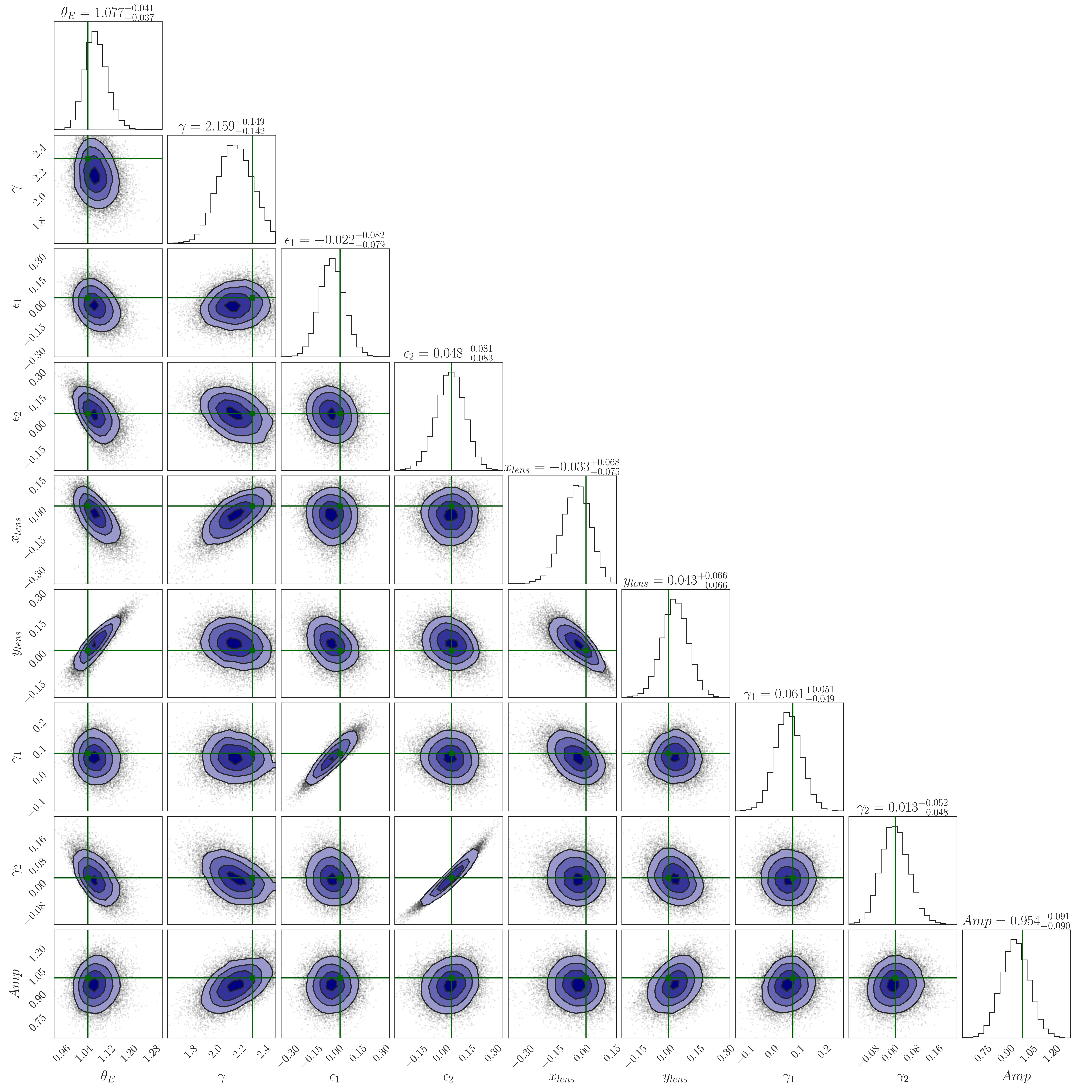}
      \begin{tikzpicture}[remember picture,overlay]
      \node at (-3.0cm,14.0cm) 
      {\includegraphics[scale = 0.40]{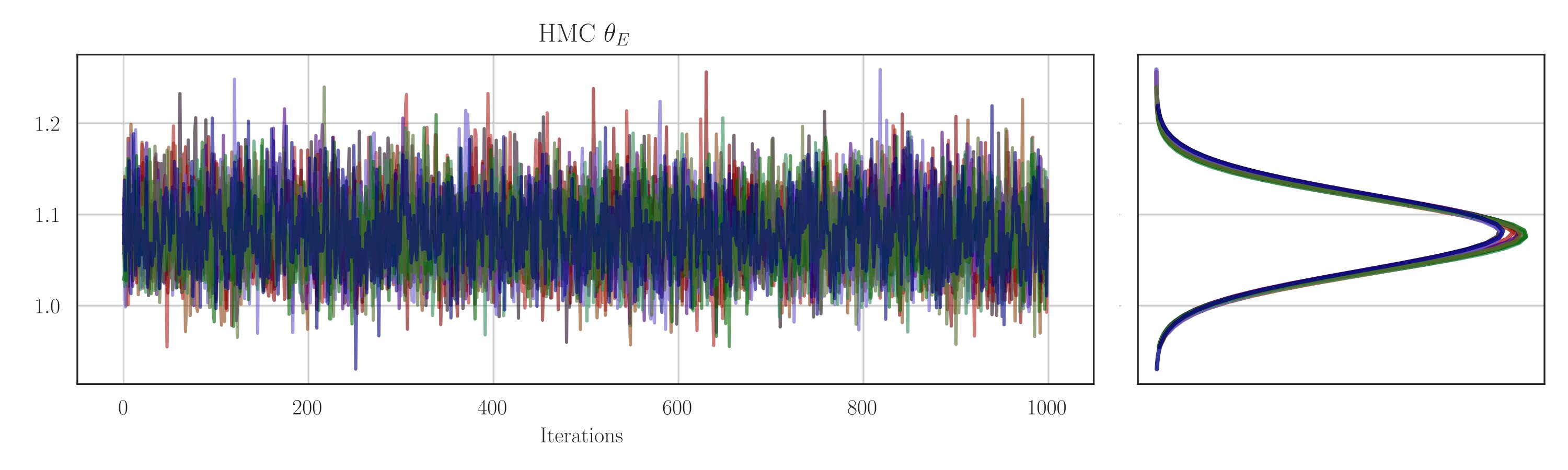}};
    \end{tikzpicture}
    \begin{tikzpicture}[remember picture,overlay]
      \node at (-3.1cm,11.0cm) 
      {\includegraphics[scale = 0.40]{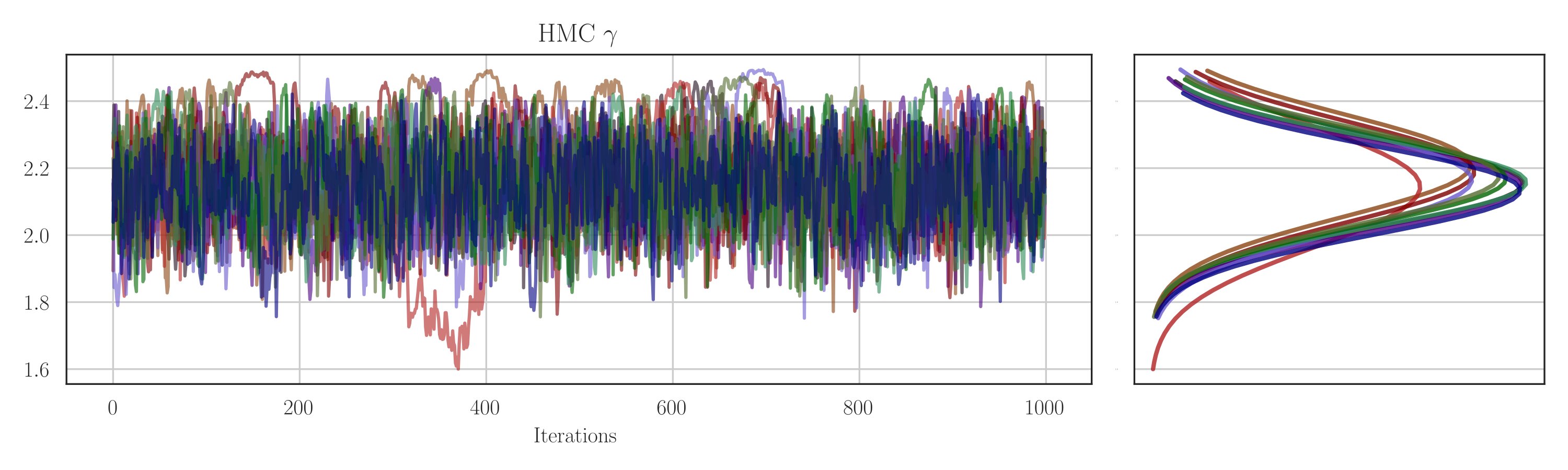}};
    \end{tikzpicture}
    \\\includegraphics[keepaspectratio=true,scale=0.40]{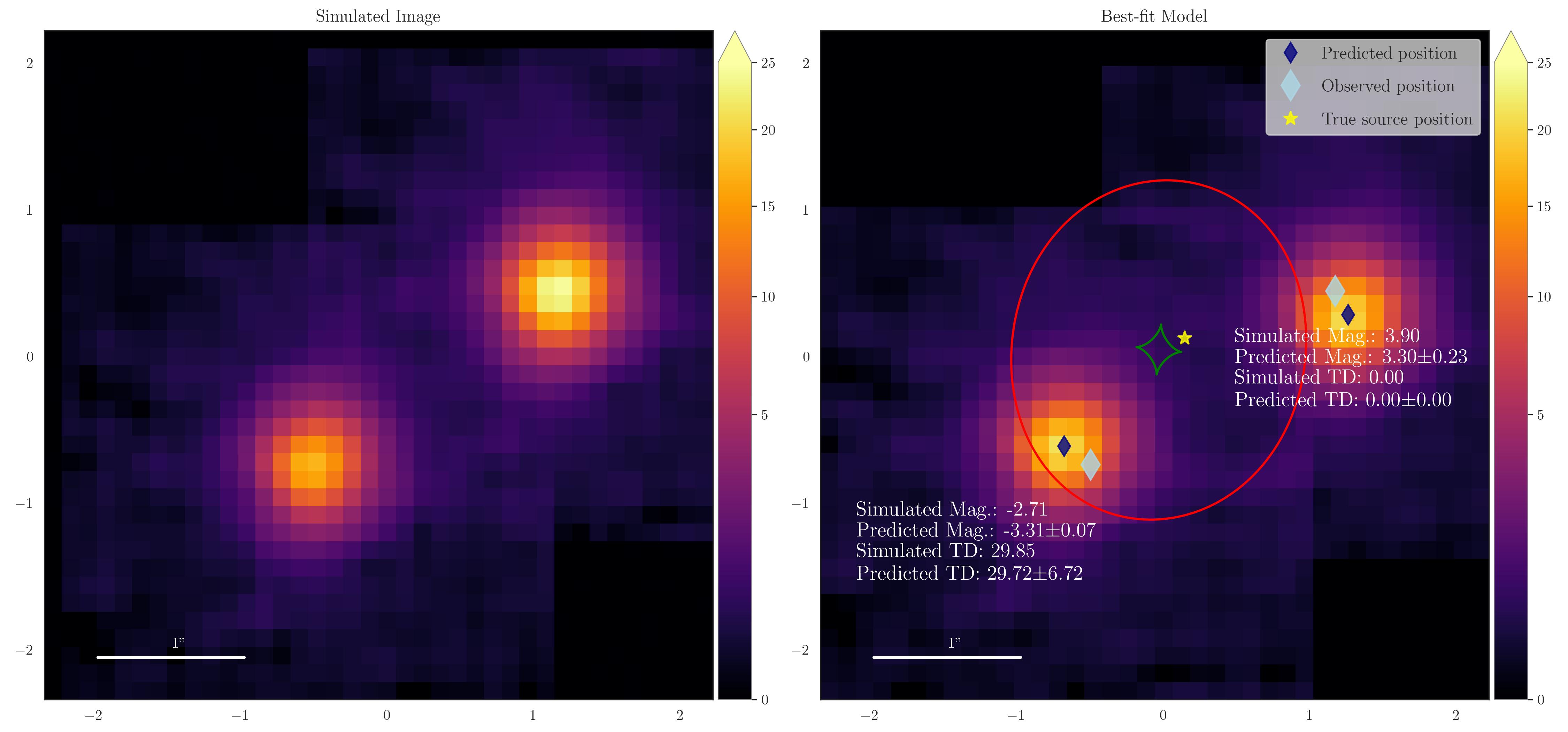}
    
    \caption{Modeling results for the double configuration. The best-fit model shows decent predictions for flux and time delay but notably less accurate positions compared to other configurations. Regardless, the sampling notably achieves $\rhat_{max} =  1.015$ and ESS $= 943-15132$, after 20 min.\ and 50 sec.\ of modeling time.}
    \label{fig:double}
\end{figure}

Each of the five archetypal configurations is determined by the location of the source relative to the caustic. 
In the cross configuration, the source is located inside the caustic. The corresponding model and corner plot are illustrated in Figure~\ref{fig:cross}.
The long-cusp and short-cusp configurations are distinguished by the proximity of the source to either a long-axis or short-axis cusp of the caustic, respectively. These are characterized by the presence of three images near the critical curve that, as the source approaches the cusp, coalesce into a single image. The key difference between the long-cusp and short-cusp lies in the parity of the isolated image: negative parity in the former configuration and positive parity in the latter. These differences in parity have implications for the arrival times of the lensed images, as the first two arriving images need to have positive parity \citep{narayan1996a}. The models and sampling results for these systems are presented in Figures~\ref{fig:long-cusp} and~\ref{fig:short-cusp}, respectively.
The modeling results for the fold configuration, with two of the lensed images merging across the critical curve, are shown in Figure~\ref{fig:fold}.
Finally, in the double configuration, the source is located outside the caustic, thus reducing the number of lensed images to two, as seen in Figure~\ref{fig:double}.
For each system, 
we use a set of ground truth parameters to simulate image positions, flux, and time-delay; which constitute all that the pipeline requires for modeling.
The prior distributions used in modeling these systems are specified in Table~\ref{tab:archetypal-prior}.

\begin{table}[h]
\caption{Prior distribution used for lens modeling of archetypal systems.}
\begin{align*}
\begin{cases}
    \hfill \theta_E & \sim \mathcal{U}(0.5, 2.0) \\[1pt]
    \hfill \gamma & \sim \mathcal{TN}\left(2, 0.25, 1.5, 2.5\right) \\[1pt]
    \hfill \epsilon_{1}, \epsilon_{2} & \sim \mathcal{N}(0, 0.1) \\[1pt]
    \hfill x, y & \sim \mathcal{N}(0, 0.1) \\[1pt]
    \hfill \gamma_{ext, 1}, \gamma_{ext, 2} & \sim \mathcal{N}(0, 0.1) \\[1pt]
    \hfill A & \sim \mathcal{N}(1, 0.1)
\end{cases} \\
\end{align*}
\label{tab:archetypal-prior}
{\small Notes---The mass model consists of EPL for the lens mass profile and external shear ($\gamma_{ext}$), with a parameter for the intrinsic flux of the \Ia, $A$, normalized to 1.0. 
$\theta_E$ is the Einstein radius in arcsec, while $\gamma$ defines the slope of the EPL profile.
$x$ and $y$ are the mass center coordinates of the lens. 
$\gamma_{\text{ext},1}$ and $\gamma_{\text{ext},2}$ are the external shear components.
The parameters $\epsilon_1$ and $\epsilon_2$ are the lens mass eccentricities.
Finally, $\mathcal{U}(a, b)$ indicates a uniform distribution with support $[a, b]$, $\mathcal{N}(\mu, \sigma)$ is a Gaussian with mean $\mu$ and standard deviation $\sigma$, and $\mathcal{TN}(\mu, \sigma; a, b)$ is a truncated Gaussian with support $[a, b]$.}
\end{table}

From Figure~\ref{fig:cross} to~\ref{fig:double}, the corresponding best-fit model shows both the ground truth and predicted values, and the corner plot shows the sampling results for all parameters of the model. In the inset, ten chains of HMC demonstrate statistical consistency within and across chains for the Einstein radius $\theta_E$ and the mass slope $\gamma$.

All models---with the exception of the double configuration---yield excellent predictions, recover nearly all parameters within $1\sigma$, and exhibit statistically significant sampling, with $\rhat \leq 1.01$ and \mbox{ESS $\gtrsim \mathcal{O}(10^3)$}
for all parameters across all simulations. They demonstrate the robustness of the pipeline within an average modeling time of 13 min.\ 49 sec.\ using four A100 GPUs on a single GPU node at Perlmutter supercomputer. 
Our modeling pipeline is capable of reliably analyzing any of these configurations, offering a powerful tool for constraining the mass parameters of lensing galaxies. 
With these five simulations, we demonstrate that our pipeline can consistently constrain the mass slope $\gamma$ using only information from the lensed point source. \ed{This highlights the effectiveness of our loss function.
In the JAX implementation, all models have achieved $\rhat \leq 1.1$ normally within seconds. By increasing the number of samples and adjusting the number of intermediate steps, it has been possible to achieve $\rhat \leq 1.01$, typically in the order of minutes.}

In the case of the double system, the available information is reduced by half, leading to a significant loss of constraining power in the loss function. Even so, we find good convergence (measured by $\hat{R}$ and ESS) and obtain accurate predictions for both positions and fluxes. Furthermore, the image parities are correctly recovered, and the predicted time delays are in excellent agreement with the simulations.
Given the abundance of doubly lensed quasars, these systems still represent a powerful probe of \ho, owing to their typically large relative time delays, which arise from highly asymmetric separation of the two images with respect to the lens center.

\subsection[H0 fitting]{\ho fitting}\label{subsec:H0}
\sbj{In this subsection we present the modeling results for the simulation of a typical lensed point source. This system provides the first fully forward model in which \ho is treated explicitly as a fitting parameter. It consists of a cross configuration with a higher-than-isothermal mass slope, $\gamma = 2.1$, and eccentricities $\epsilon_1 = 0.1$ and $\epsilon_2 = -0.05$, corresponding to an axis-ratio of $q = 0.80$. An Einstein radius of $\theta_E = 2.15''$, together with the mild asymmetry introduced by the ellipticity, yields time-delays on the order of days.
We employ a ground truth of $\ho = 70$~km$\,\mathrm{s}^{-1}$/Mpc, and a uniform prior distribution $\mathcal{U}(0,150)$---favoring no current measurement of the Hubble constant. We assume a $\Lambda$CDM cosmology, with the lens and source placed at \snz's redshifts $z_d = 0.2262$ and $z_s = 0.3544$, respectively. For the lensing parameters, we use the same prior distributions as those of the archetypal systems, found in Table~\ref{tab:archetypal-prior}, except for the Einstein radius. Due to the larger size of the system, we assign a uniform prior $\theta_E \sim \mathcal{U}(1.5,3.0)$.}


\begin{figure}[ht]
    \centering
    \includegraphics[keepaspectratio=true,scale=0.28]{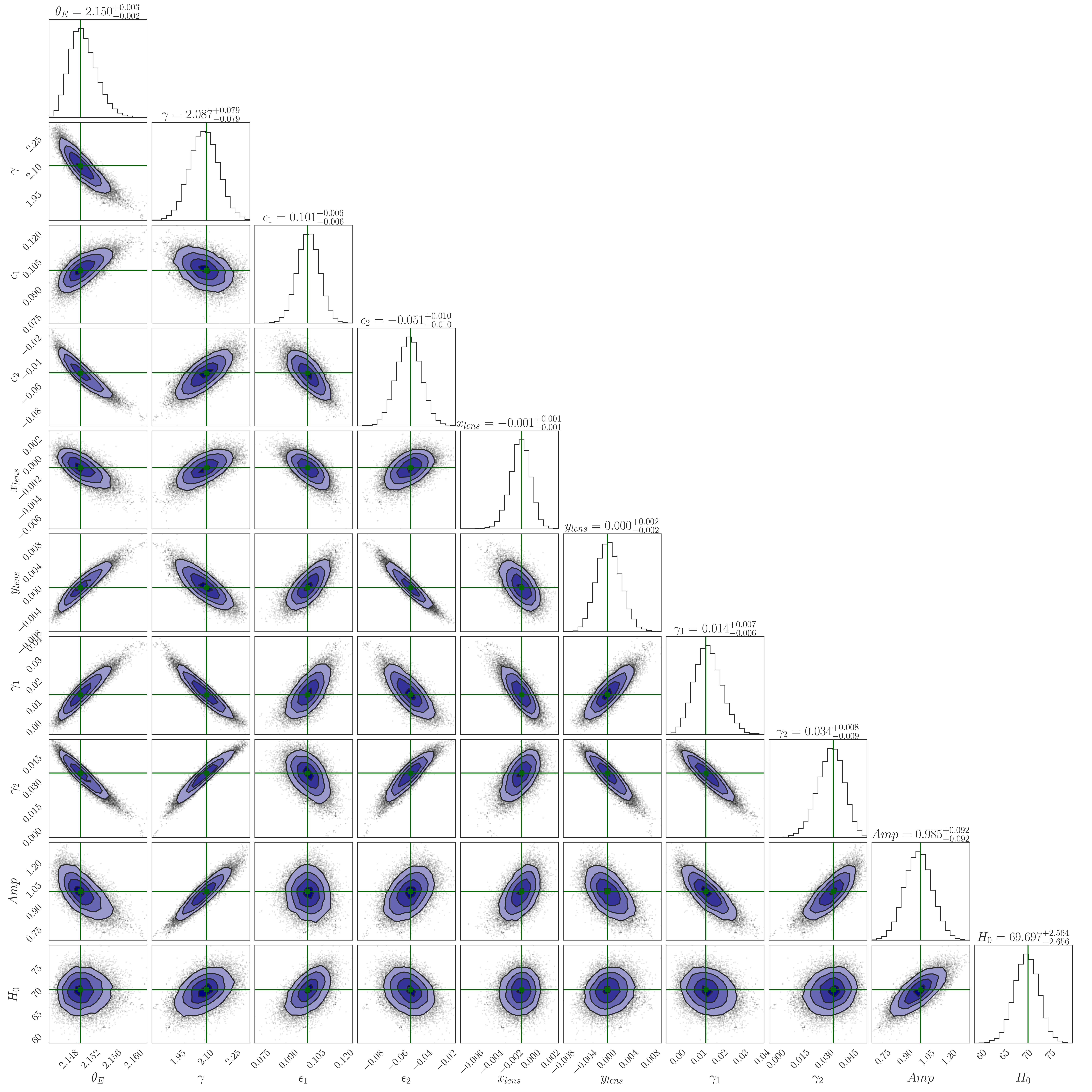}
      \begin{tikzpicture}[remember picture,overlay]
      \node at (-3.8cm,14.0cm) 
      {\includegraphics[scale = 0.37]{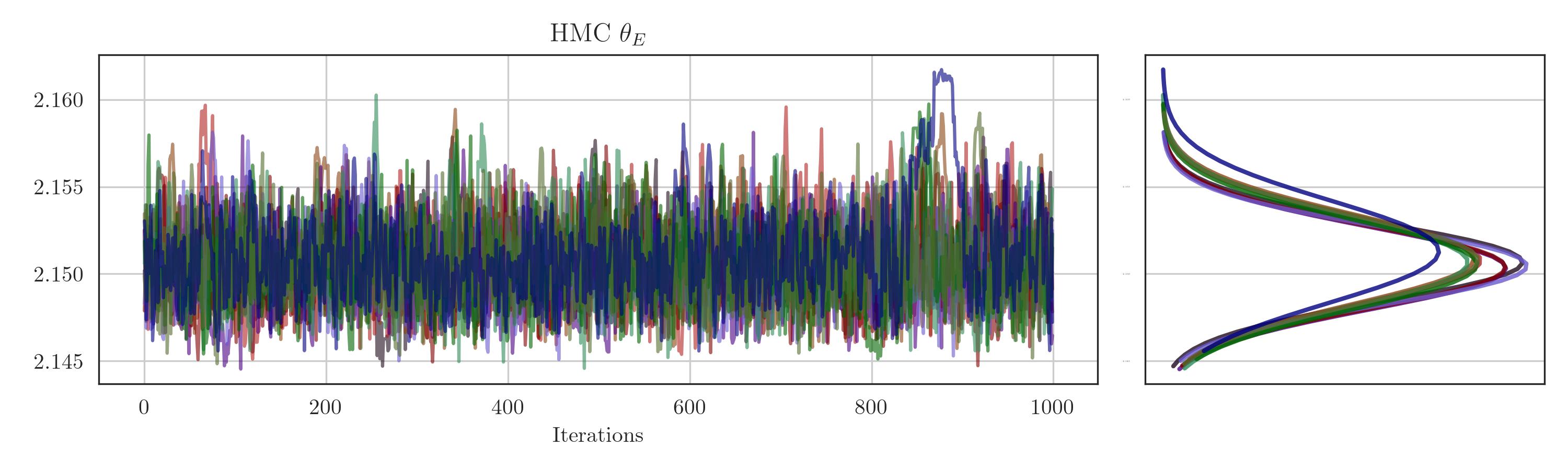}};
    \end{tikzpicture}
    \begin{tikzpicture}[remember picture,overlay]
      \node at (-3.9cm,11.2cm) 
      {\includegraphics[scale = 0.37]{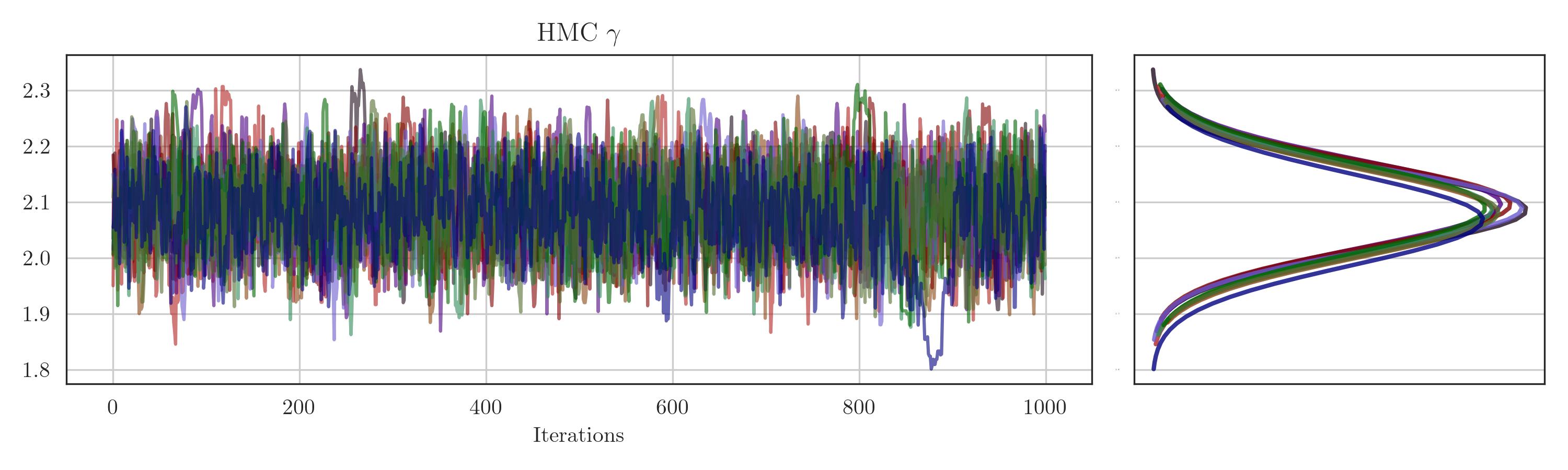}};
    \end{tikzpicture}
    \\\includegraphics[keepaspectratio=true,scale=0.37]{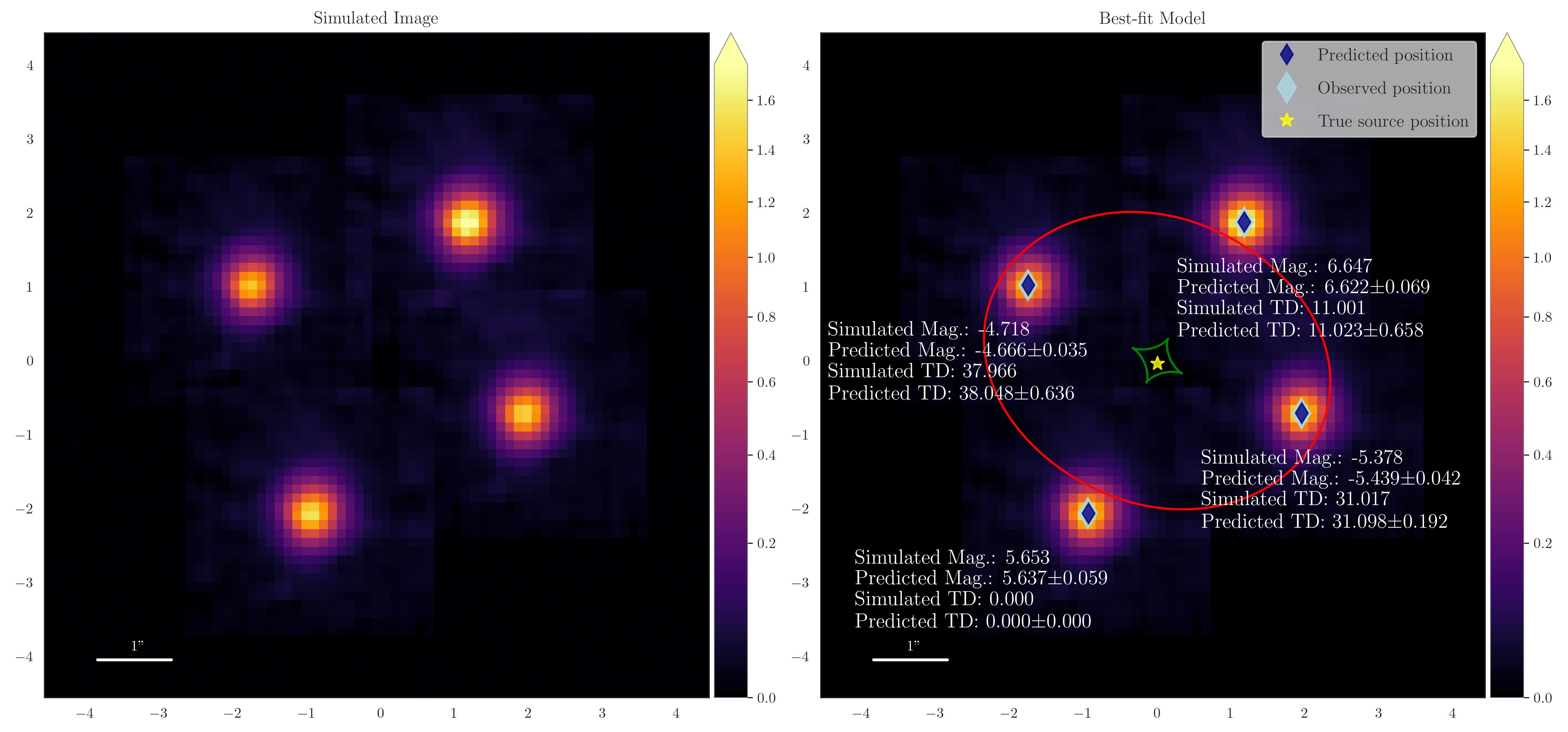}
    
    \caption{\sbj{Modeling results of a system fitting for \ho. The sampling achieves $\rhat_{max} =  1.007$, \mbox{ESS $= 4289-17630$}, 
    within 5 min.\ and 21 sec.\ of modeling time. Importantly, the lower-right corner of the sampling results shows the marginal distribution of \ho, with a statistical uncertainty of $3.6\%$.}
    }
    \label{fig:H0-system}
\end{figure}

\begin{figure}
    \centering
    \includegraphics[scale = 0.7]{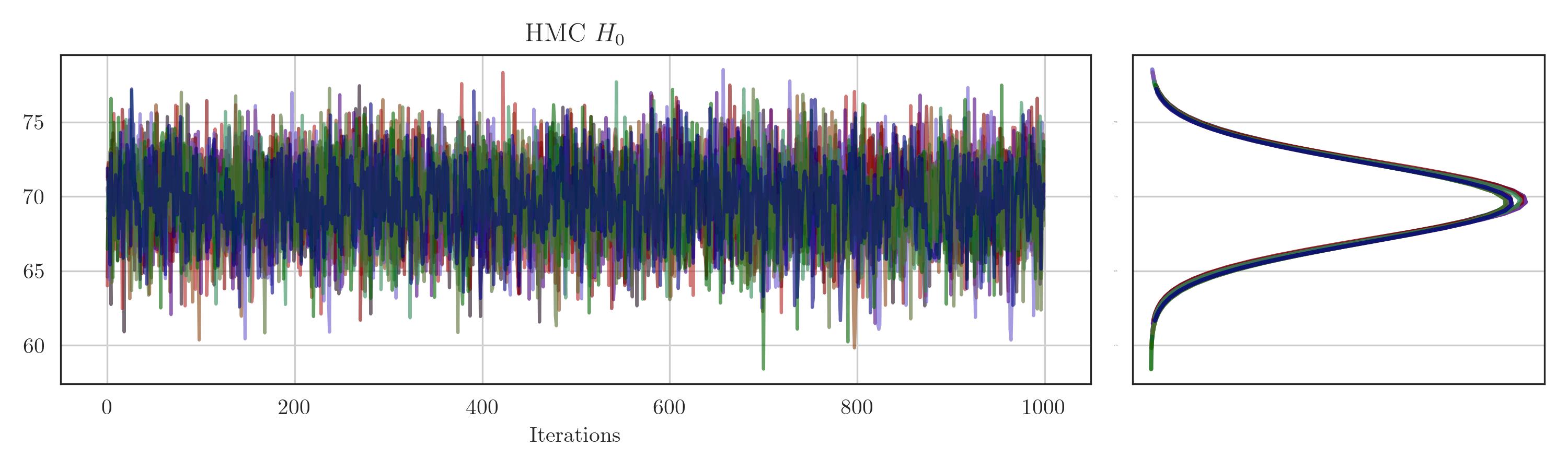}
    \caption{\sbj{HMC sampling traces for \ho. The chains exhibit good mixing and stable convergence behavior, with a $3.6\%$ standard deviation.}}
    \label{fig:chains-H0}
\end{figure}

Figure~\ref{fig:H0-system} shows our highly accurate predictions for image positions, fluxes and time delays, with low uncertainties. Additionally, all parameters are recovered within $1\sigma$---including \ho, \sbj{with a standard deviation of $\sim 3.6\%$} (see cornerplot in Figure~\ref{fig:H0-system} and traces in Figure~\ref{fig:chains-H0}). 
We thus have demonstrated the effectiveness of our pipeline in constraining \ho, requiring as few data points as 16, corresponding to a \textit{single} lensed SN~Ia with relevant time delay observations. 
\sbj{Even though we report a very standard configuration, we found that systems with larger Einstein radii yield smaller statistical uncertainties on \ho, consistent with the fact that time delays scale with $\theta_E$. In addition, lenses with higher ellipticity introduce greater asymmetry in the image configuration, thereby increasing the relative differences in the time delays between images. We observe that this higher asymmetry also leads to tighter constraints on \ho. Hence, systems such as \snz---were it to have a larger Einstein radius---are capable of providing very stringent measurements of \ho.}

The leading systematic uncertainty in inferring \ho from strongly lensed SNe is the mass-sheet degeneracy \citep[MSD;][]{Schneider2013a}. If the lensed SN is a Type~Ia, its standardizable luminosity provides an absolute magnification measurement, which directly constrains the mass normalization and mitigates the MSD \citep[e.g.,][]{huber2019}. 
This ingredient is already incorporated into the modeling pipeline used in this work. 
While the  current implementation does not include a mass-sheet component, its incorporation would be straightforward. We leave this extension of our pipeline to future work.
In parallel, spatially resolved stellar kinematics provide an independent measurement of the gravitational potential, anchoring the enclosed mass profile and reducing systematics due to MSD \citep[e.g.,][]{yildirim2023}. It is possible to combine lens modeling and stellar-dynamics modeling in a unified inference framework \citep[e.g.,][]{wang2025}. Although its name emphasizes lensing, \gigal is in fact a general, differentiable Bayesian modeling framework and is readily extendable beyond lensing applications. We leave a full joint lensing-dynamics analysis to future work.

%% file: real-systems.tex
In this section, we present the application of our method to two real systems: SN~Ia 2022qmx (also known as ``SN Zwicky") and Type Ia \geu.
Our results for these systems not only demonstrate the effectiveness of our pipeline in real data, but they also serve as a comparison with previous models.

\subsection{SN iPTF16geu}\label{sec:16geu}
\input{16geu.tex}

\subsection{SN Zwicky}\label{sec:snzwicky}
\input{snzwicky.tex}

%% file: 16geu.tex
\begin{figure}
\centering
  \includegraphics[width=.4\linewidth]{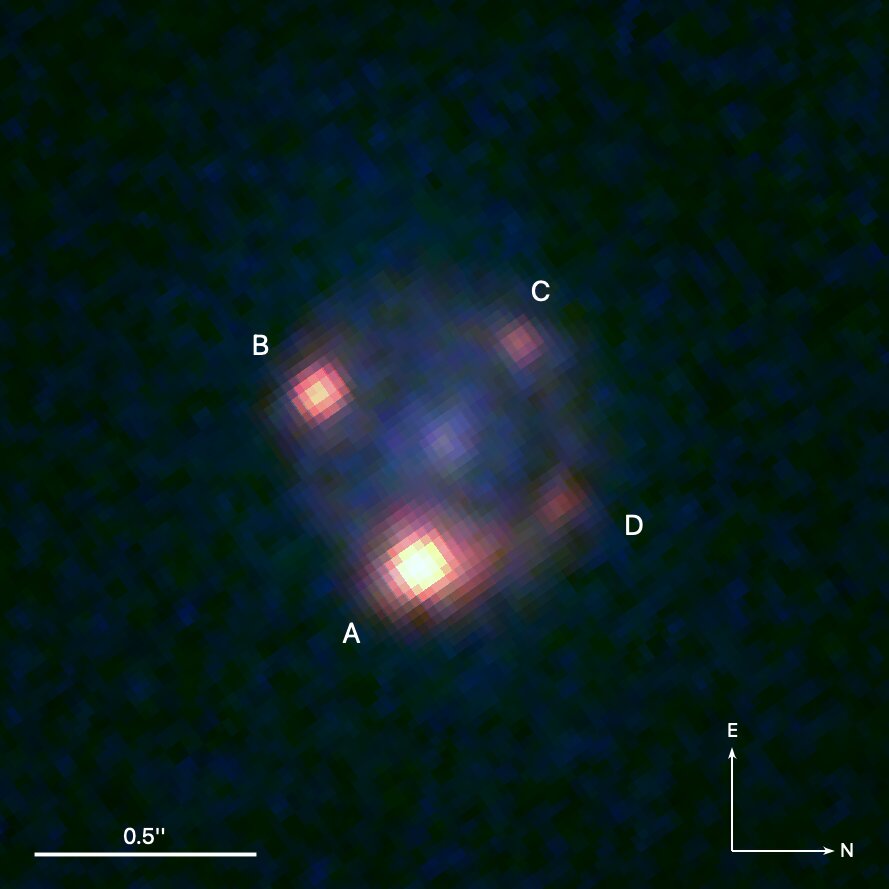}
\caption{RGB image of \geu generated from \hst GO--14862 (PI: A.~Goobar) observations (F814W, F625W, and F475W). The images of the lensed source, forming an Einstein Cross-like pattern, are indicated with A, B, C, and D.}
\label{fig:16geu-obs}
\end{figure}

The strongly lensed \geu was discovered in 2016 through the intermediate Palomar Transient Factory (iPTF) \citep[][G17]{goobar2017a} at a redshift of $z_s = 0.409$ and with a lens galaxy at $z_d = 0.2163$. It consists of a galaxy-scale strongly lensed SN~Ia, being the first of only three such systems reported to date, with \snz being the second. In this system, four images forming an Einstein cross pattern (Figure~\ref{fig:16geu-obs}) are approximately $0.3''$ from the center of the lensing galaxy. 
G17 provided a preliminary lens model of \geu, followed by a second model by \citet[][M17]{More2017a}.
\citet{Dhawan2019a} updated the image positions, magnifications, and time delays using high resolution \HST (\hst) and Keck Observatory AO data, once the reference image was available.
\citet[][M20]{Mortsell2020a} modeled the updated imaging data, \xh{using the lensed host galaxy}, and combined it with stellar velocity dispersion. 
 However, the uncertainties in the measured time delays are comparable in magnitude to their values, limiting their effectiveness for constraining both the mass model and \ho.
 Their analysis also examined the effects of microlensing on the model, particularly in response to flux ratio anomalies identified by the lens model.

In this work, we apply our methodology to this system, and present a fully forward model that utilizes \emph{only data about the point source}---its image positions and observed magnifications from \citet{Dhawan2019a}, who leveraged the standardizability of \Ia to report magnifications instead of fluxes.
\sbj{We employ the same prior distributions as those used in the simulations (see Table~\ref{tab:archetypal-prior}) for all parameters, with the exception of the Einstein radius. For this system, the Einstein radius prior is replaced with a shifted uniform distribution to account for its smaller size, $\mathcal{U}(0, 0.5)$.}

Our model comprises a mass model for the lens (EPL and external shear), as well as the intrinsic amplitude of the SN, for a total of 9 parameters\footnote{\sbj{As a reminder, unlike \gigal for extended sources, which has the unlensed source position coordinates as model parameters, this pipeline does not. Instead, the image positions are traced back to the source plane using the lensing-potential parameters, and we minimize the separation between these delensed positions.}}.
The modeling results can be found in Table~\ref{tab:16geu-best-fit} along with previously reported results.
In M17, two mass models are used, a Singular Isothermal Ellipsoid (SIE) and EPL + External Shear. In the latter case, they determined the mass slope to be $\gamma \simeq 2.1$.
When the final reference image became available, the combined analysis of lens modeling and velocity dispersion from M20 yielded $\gma \simeq 1.8$.
Our best-fit parameters are in very good agreement with those found in M20.
In particular, our best-fit mass slope, $\gma \simeq 1.9$, is within $1\sigma$ of their value.
However, while M20 determined the mass slope using velocity dispersion measurements and subsequently fixed this value to conduct lens modeling, our analysis independently arrives at \gma through point source information only via a fully forward modeling approach; we did not use velocity dispersion information (although in a future extension of our methodology, it is possible to also incorporate velocity dispersion in the modeling).
We also report similar flux ratio anomalies across all images, which were attributed to microlensing effects (see M20).
Our results suggest that a smooth lens model alone cannot account for the observed fluxes, in agreement with the conclusions from M17 and M20.

\xh{We note that in our model, the density slope $\gamma$ is larger by $\Delta \gamma \simeq 0.1$, resulting in slightly rounder critical curves and a moderately larger inner caustic. We also infer a slightly higher external shear amplitude.
These differences should be interpreted in the context of the modeling methodology. Our analysis employs full forward modeling of the lensed point source images, whereas M20 did not explicitly include external shear as a free model component (see their Equation~30 and the text around it). 
Full forward modeling allows these contributions to be treated self-consistently within the likelihood, which can lead to modest shifts in best-fit parameters even when the overall lensing configuration remains similar.
Despite these parameter-level differences, the predicted image positions relative to the tangential critical curve are nearly identical between the two models. Consequently, the predicted ordering of the time delays is also unchanged.}

\begin{figure}[H]
  \centering
  \vspace{0.2cm}
  \includegraphics[scale = 0.33]{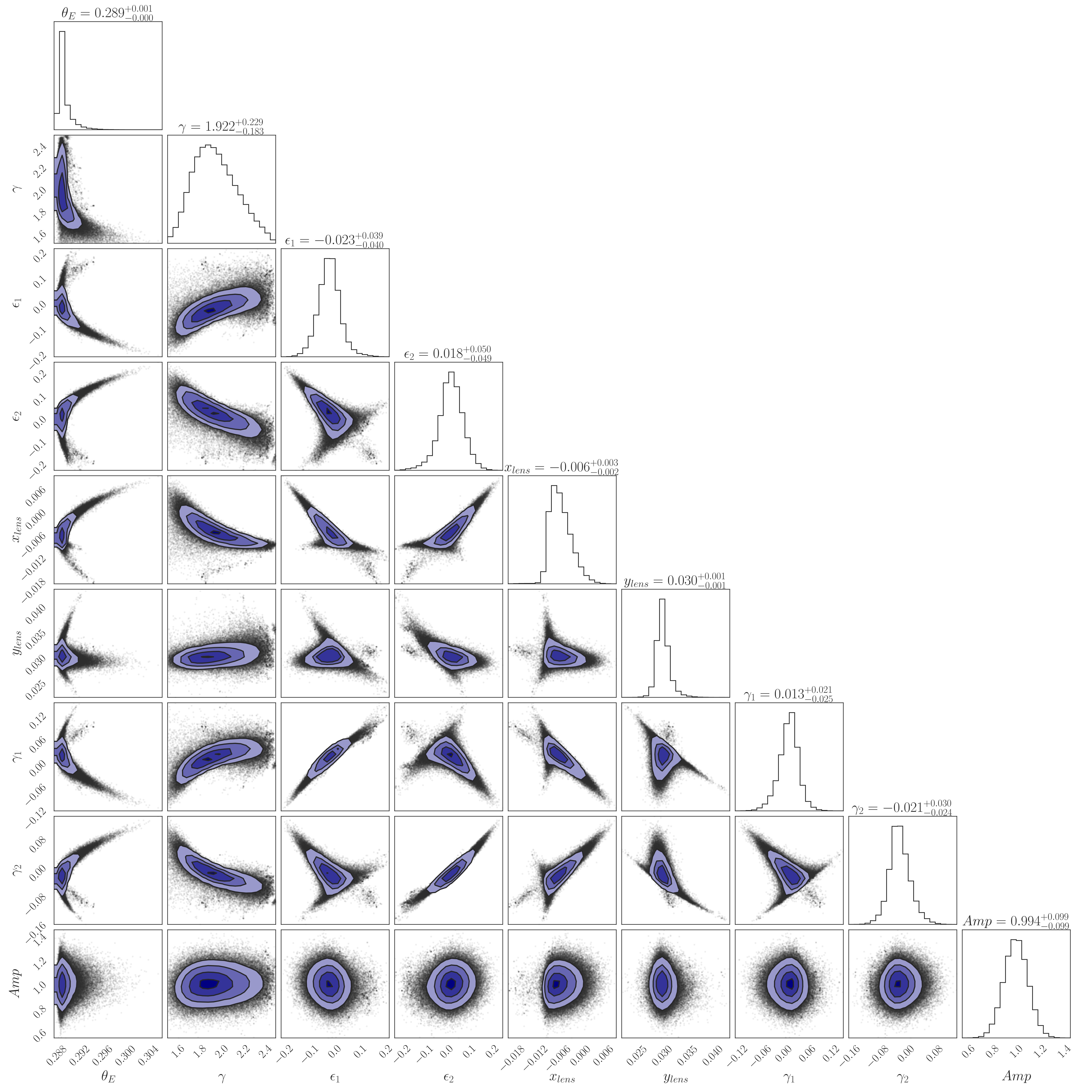}
  \begin{tikzpicture}[remember picture,overlay]
      \node at (-3.7cm,15.5cm) 
      {\includegraphics[scale = 0.40]{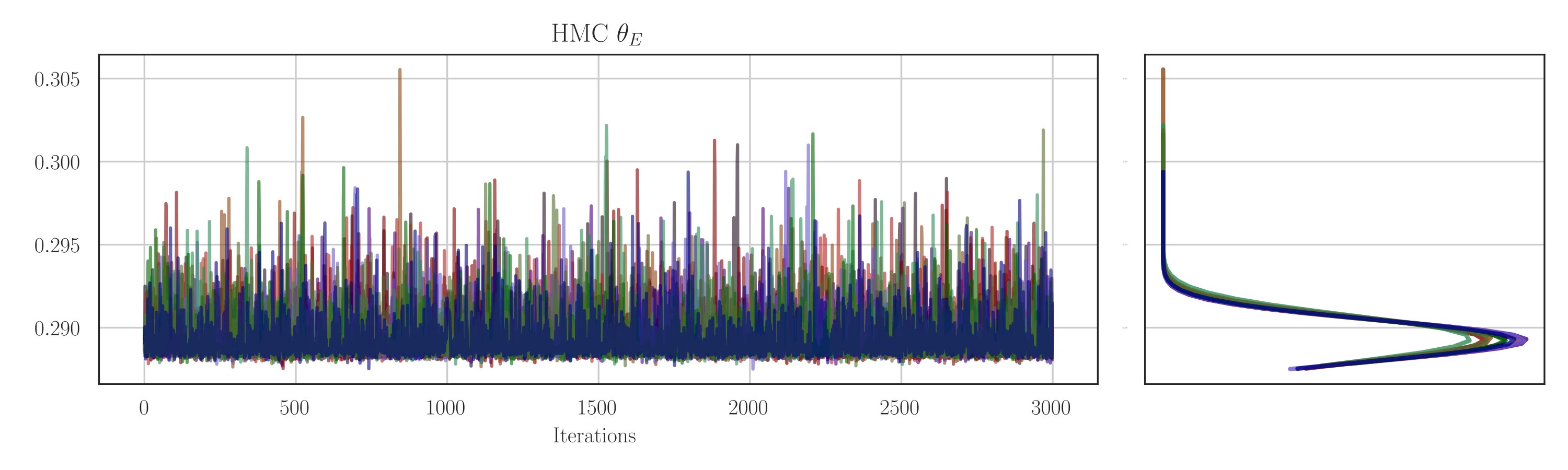}};
    \end{tikzpicture}
  \begin{tikzpicture}[remember picture,overlay]
      \node at (-3.92cm,12.5cm) 
      {\includegraphics[scale = 0.40]{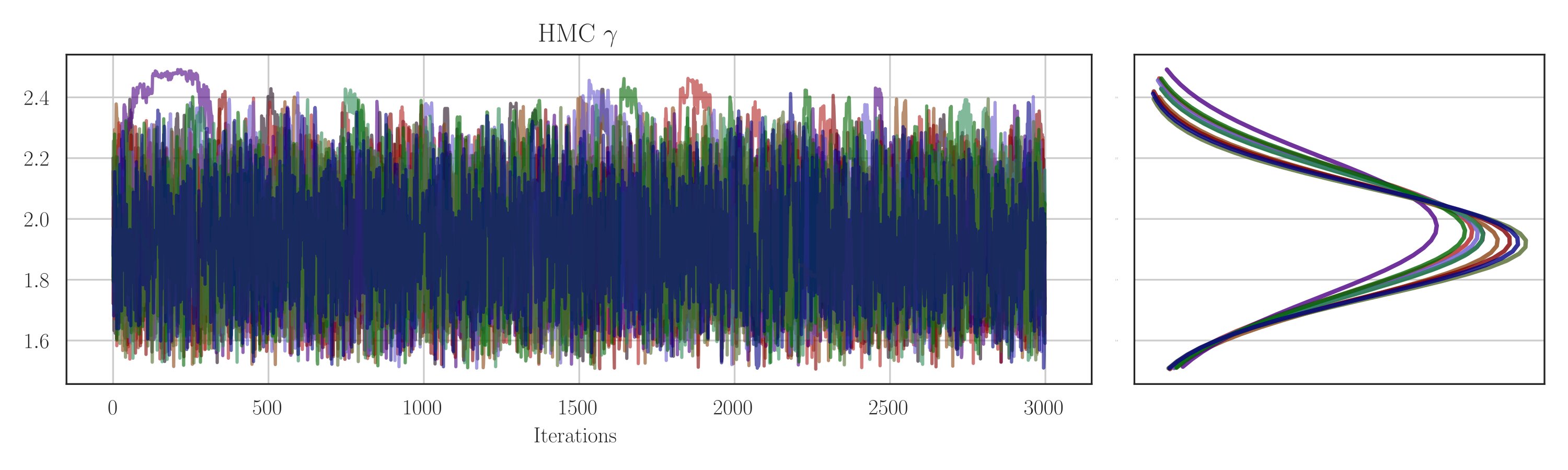}};
    \end{tikzpicture}
  \begin{tikzpicture}[remember picture,overlay]
      \node at (-2.7cm,10cm) 
      {\includegraphics[scale = 0.28]{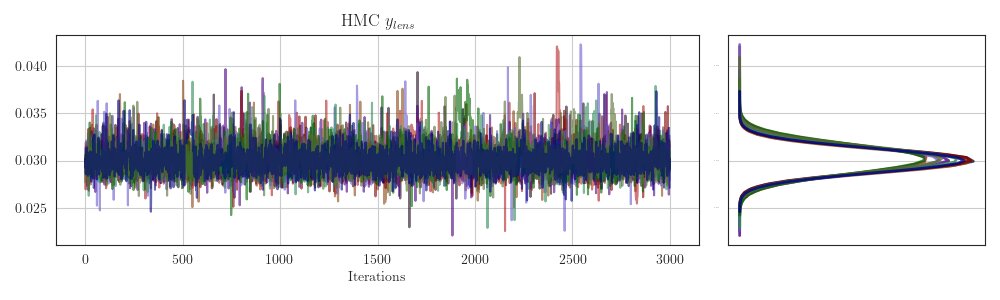}};
    \end{tikzpicture}
  \begin{tikzpicture}[remember picture,overlay]
      \node at (-2.93cm,7.7cm) 
      {\includegraphics[scale = 0.28]{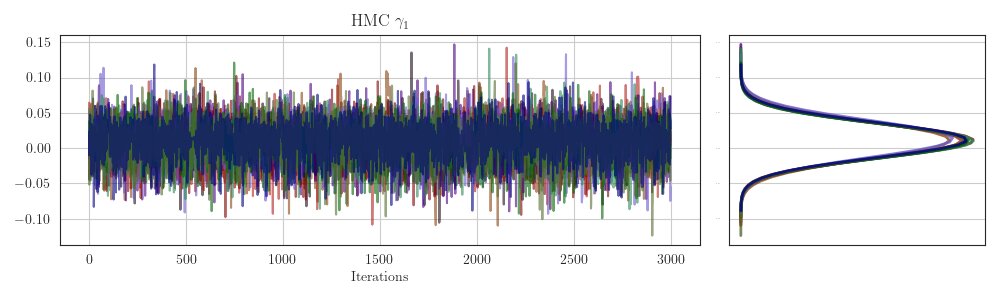}};
    \end{tikzpicture}
  \caption{\geu. The corner plot shows our sampling results for every parameter in our model, achieving $\rhat_{max} = 1.003$, and ESS $= 6153-132052$ within 57 min.\ and 19 sec.\ of modeling time. In the inset, ten chains of HMC demonstrate statistical consistency within the sampling of the Einstein radius, $\theta_E$, the mass slope, \gma, $y_{lens}$, and $\gamma_1$. 
  }
  \label{fig:corner-JAX-16geu}
\end{figure}

In Figure~\ref{fig:16geu-best-model}, we present the predicted and observed values alongside an image of our best-fit model. 
The sampling results are displayed in Figure~\ref{fig:corner-JAX-16geu}.
With 12 available data points for this system (eight positional quantities ($x$, $y$) corresponding to the four images and four observed magnifications) and nine parameters in our model, we achieve excellent convergence as demonstrated by $\rhat_{\text{max}} = 1.003$, with a total modeling time of 57 minutes and 19 seconds.
Once again, if we merely wanted to reach the point of producing a “presentable” corner plot, or just achieving 
$\rhat < 1.1$, the modeling time would be an order of magnitude shorter.

\begin{deluxetable*}{l|ccccccccc}
\tabletypesize{\scriptsize}
\tablecaption{Best-fit parameters for \geu compared with previous results \citep[][]{goobar2017a, More2017a, Mortsell2020a}. From \citet{More2017a}, we compare only their results from their EPL model.
All parameters are defined in Table~\ref{tab:archetypal-prior}. 
\sbj{Unreported values are shown as ``$-$''. The label ``Not fit'' indicates parameters that were excluded from the model, and were fixed to the value shown in parenthesis.}
\label{tab:16geu-best-fit}}
\renewcommand{\arraystretch}{1.4}
\tablehead{
    \colhead{Team} &
    \colhead{$\theta_E$} &
    \colhead{$\gamma$} &
    \colhead{$\epsilon_1$} &
    \colhead{$\epsilon_2$} &
    \colhead{$x$} &
    \colhead{$y$} &
    \colhead{$\gamma_{ext, 1}$} &
    \colhead{$\gamma_{ext, 2}$} &
    \colhead{$A$}
}
\startdata
This work &
$0.289_{-0.0001}^{+0.0010}$ & 
$1.922_{-0.183}^{+0.229}$ & 
$-0.023_{-0.040}^{+0.039}$ & 
$0.018_{-0.049}^{+0.050}$ & 
$-0.006_{-0.002}^{+0.003}$ & 
$0.030_{-0.001}^{+0.001}$ & 
$0.013_{-0.025}^{+0.021}$ & 
$-0.021_{-0.024}^{+0.030}$ & 
$0.994_{-0.099}^{+0.099}$
\\
Goobar+&
$0.287 \pm 0.005$& 
Not fit ($2.0$)&	
$-0.11\pm 0.06$&
$0.12 \pm 0.07$  & 
$-$&
$-$ &
Not fit ($0.0$)&	
Not fit ($0.0$)&	
$-$
\\
More+&
$0.30\pm 0.01$& 
$2.1 \pm 0.1$&	
$-0.15_{-0.02}^{+0.04}$&
$0.14_{-0.02}^{+0.04}$  & 
$-$&
$-$ &
$-0.0186_{-0.009}^{+0.031}$&	
$0.0075_{-0.004}^{+0.015}$&
$-$
\\
M\"{o}rtsell+&
$0.292 \pm 0.001$& 
$1.8^\ast$&	
$-0.042\pm 0.004$&
$0.048\pm 0.004$  & 
$-$&
$-$ &
$\sim 10^{-4}$&	
$\sim 10^{-4}$&
$-$
\\
\enddata
\tablenotetext{*}{Value found using a combination of dynamics and lens modeling.}
\end{deluxetable*}

\begin{minipage}{\linewidth}
\vspace{5pt}
\makebox[\linewidth]{
  \includegraphics[keepaspectratio=true,scale=0.6]{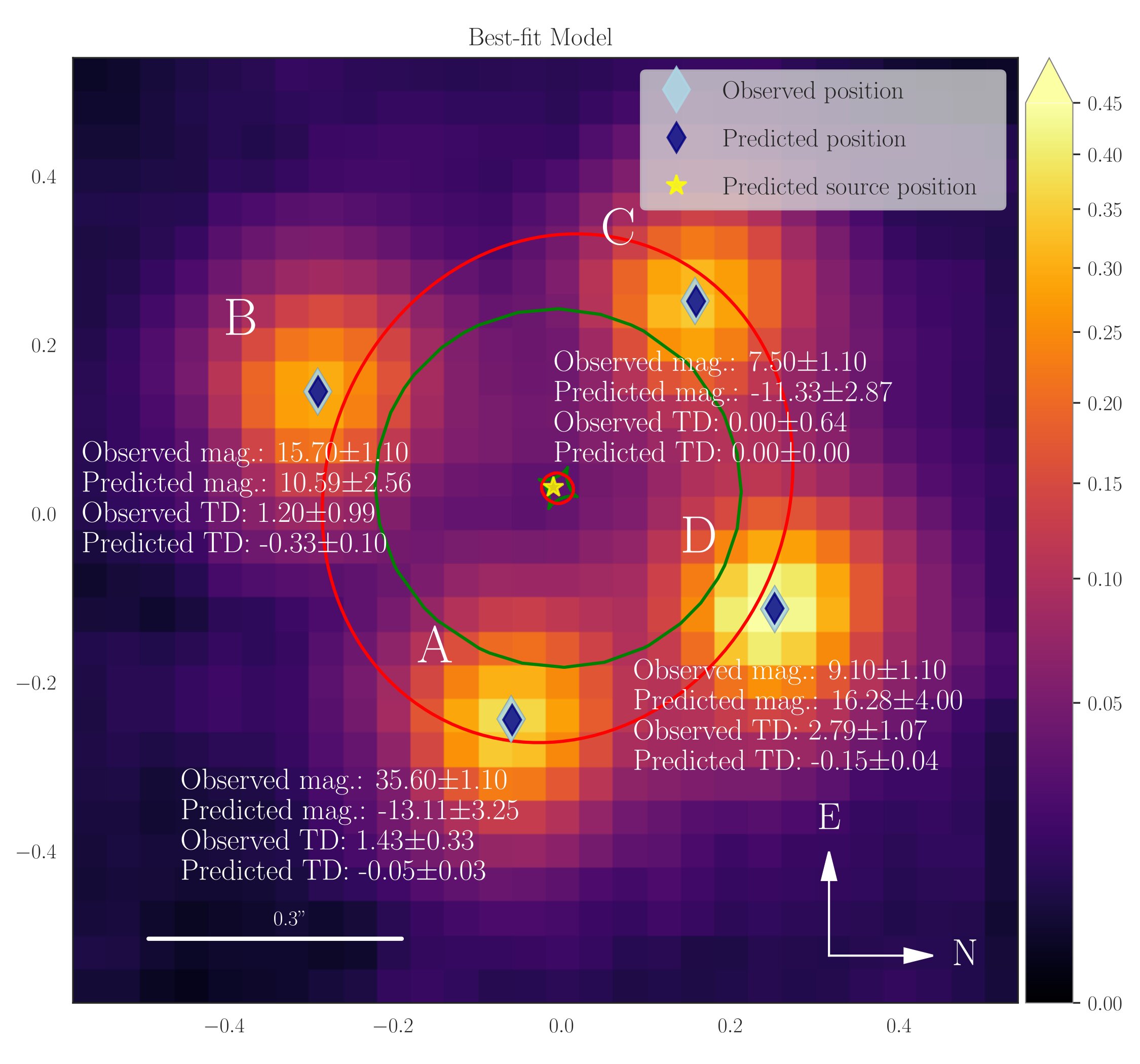}
  }
\captionof{figure}{Best-fit model for \geu. For each image, observed and predicted time delay and magnification are reported. In addition, observed positions are shown in light blue and predicted positions in dark blue. The critical curve is drawn is red while in green is the caustic. 
The predicted unlensed source position is shown as a yellow star.
The orientation and labeling match those of Figure~\ref{fig:16geu-obs}.}
\label{fig:16geu-best-model}
\vspace{10pt}
\end{minipage}

%% file: snzwicky.tex
\begin{figure}[ht]
\centering
  \includegraphics[width=.4\linewidth]{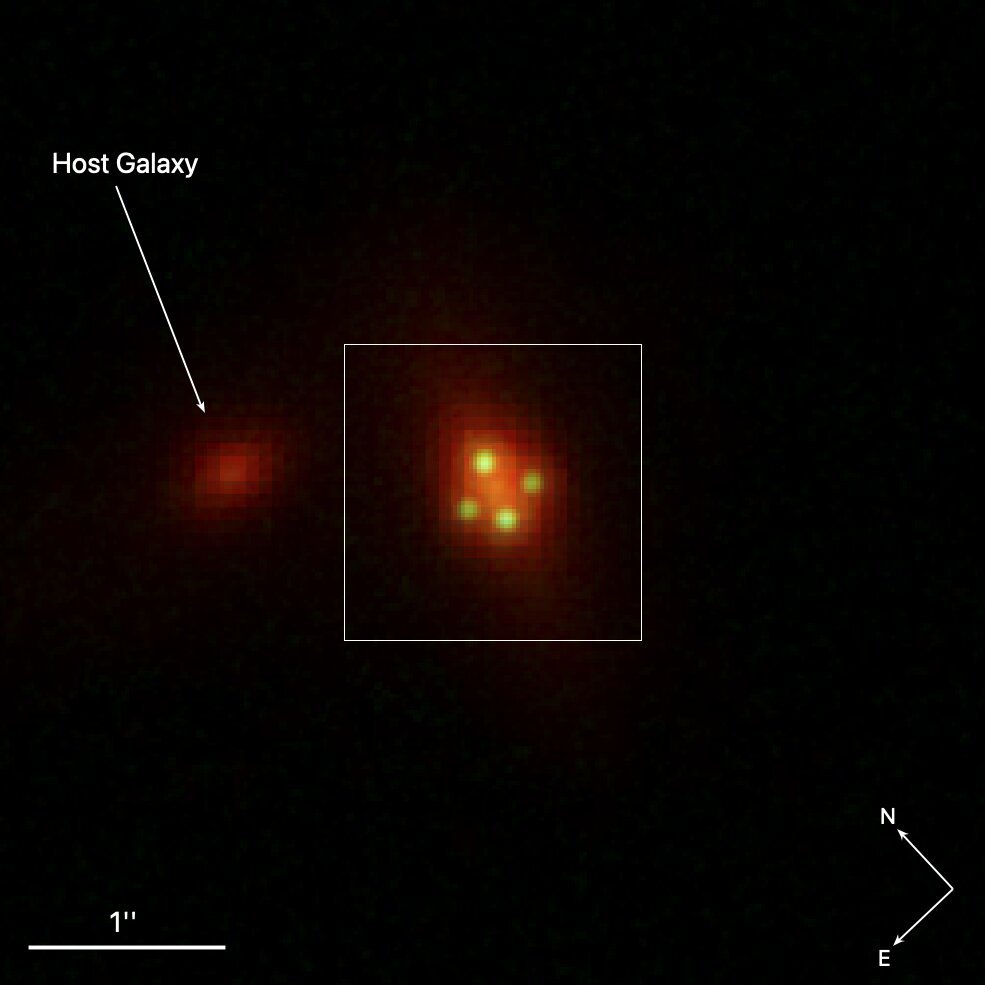}
  \includegraphics[width=.4\linewidth]{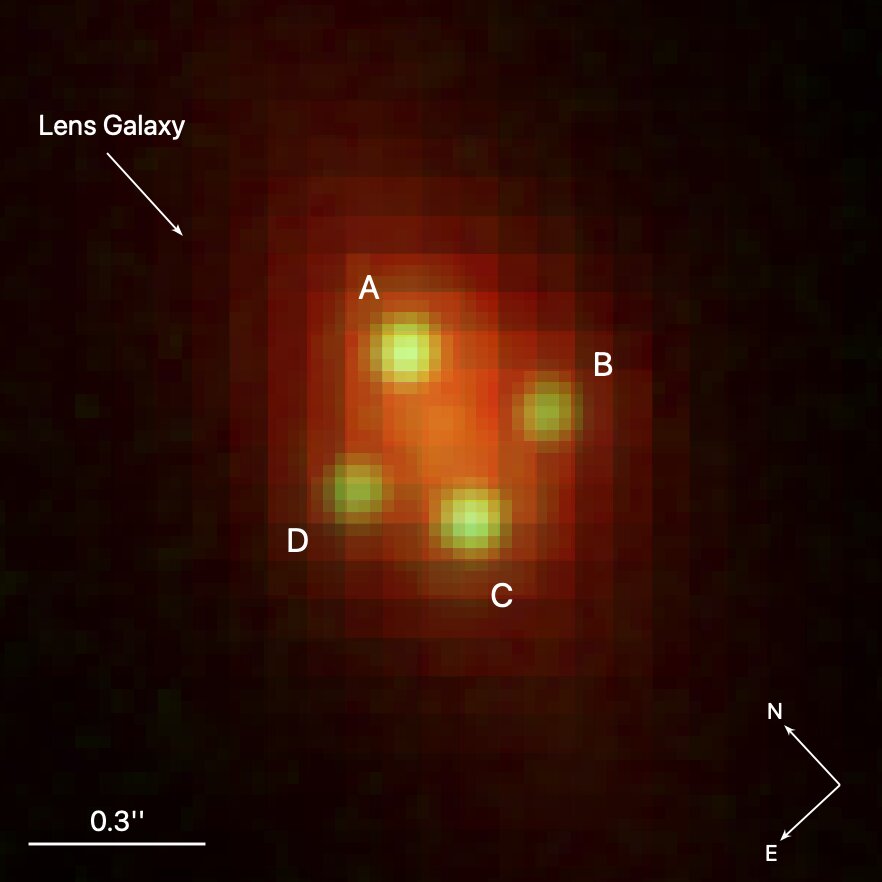}
\caption{\textit{Left panel:} RGB image of \snz, constructed from \hst GO-16264 (PI: J.~Pierel) observations (F160W, F814W, and F625W). A white arrow indicates the center of SN host galaxy. \textit{Right panel:} Zoomed-in cutout (corresponding to the white box in the left panel) showing the four lensed images, indicated as A, B, C, and D, and the lensing galaxy.
}
\label{fig:snz-obs}
\end{figure}

This system was discovered by the Zwicky Transient Facility \citep[][G23]{goobar2023a}, with a SN redshift of $\zs = 0.3544$ and a lensing galaxy redshift of $\zd = 0.2262$ obtained from spectroscopic observations from the Keck Observatory, Hobby–Eberly Telescope,
and the Very Large Telescope (VLT).
\hst observations were later reported by \citet[][P23]{pierel2023a}.
\citet[][L24]{larison2024a} further followed up with improved time delays and magnifications, which we use for our lens model.
In this system, four resolved SN images appeared to form an Einstein cross pattern around $0.17''$ from the center of the lensing galaxy.
Its host galaxy was hardly visible in initial discovery images, but its nucleus was later located $1.4''$ to the northeast of the lens (see Figure~\ref{fig:snz-obs}).

We employ our modeling pipeline using all point source information: lensed image positions, magnifications, and time delay observations from L24. Although time delay measurements were first reported in G23 and P23 (Table~6), $\lesssim0.5$~days, the uncertainties are an order of magnitude larger than the measurements themselves.
In L24, the uncertainties are significantly reduced, but they remain at the $\sim 100\%$ level. 
As discussed in \S\,\ref{sec:TD-loss}, while the time delay loss term enhances accuracy and convergence of our model, it does not serve as a critical constraint for either the mass model or \ho.

Our mass model consists of an EPL with external shear. We also fit for the amplitude of the SN, for a total of nine parameters. Even though the number of data points available is low (16 in total: eight positional quantities, ($x$, $y$) for each of the four images, four observed magnifications and four observed time delays), we are able to model the observations very accurately with only a smooth model.
\sbj{The prior distributions for all parameters can be found in Table~\ref{tab:archetypal-prior}, with the exception of the Einstein radius, for which we adopt a shifted uniform distribution $\mathcal{U}(0, 0.5)$.}

\begin{deluxetable*}{l|ccccccccc}
\tabletypesize{\scriptsize}
\tablecaption{Our best-fit parameters for SN~Zwicky compared with previous results \citep{goobar2023a, pierel2023a}. All parameters are defined in Table~\ref{tab:archetypal-prior}. Only models from \citet{pierel2023a} that report a mass slope are included in this Table (GLEE, SALT+LS, and LS2).
\sbj{Unreported values are shown as ``$-$''. The label ``Not fit'' indicates parameters that were excluded from the corresponding model, and were fixed to a default value.}
\label{tab:snz-best-fit}}
\renewcommand{\arraystretch}{1.4}
\tablehead{
    \colhead{Team} &
    \colhead{$\theta_E$} &
    \colhead{$\gamma$} &
    \colhead{$\epsilon_1$} &
    \colhead{$\epsilon_2$} &
    \colhead{$x$} &
    \colhead{$y$} &
    \colhead{$\gamma_{ext, 1}$} &
    \colhead{$\gamma_{ext, 2}$} &
    \colhead{$A$}
}
\startdata
This work &
$0.173_{-0.002}^{+0.002}$ & 
$2.109_{-0.122}^{+0.140}$ & 
$0.310_{-0.014}^{+0.015}$ & 
$0.326_{-0.015}^{+0.016}$ & 
$-0.002_{-0.001}^{+0.001}$ & 
$-0.013_{-0.001}^{+0.001}$ & 
$0.149_{-0.015}^{+0.018}$ & 
$0.155_{-0.016}^{+0.020}$ & 
$0.925_{-0.099}^{+0.099}$
\\
Goobar+ &
$0.1670\pm 0.0006$& 
2.0 (Not fit)&	
$-0.1554\pm 0.0055$&
$0.1412\pm 0.0050$  & 
$-$&
$-$ &
0.0 (Not fit)&	
0.0 (Not fit)&
$-$
\\
GLEE&
$0.168_{-0.004}^{+0.005}$& 
$2.04_{-0.22}^{+0.14}$&	
$0.27_{-0.10}^{+0.11}$&
$-0.24_{-0.09}^{+0.10}$  & 
$-$&
$-$ &
$-0.005_{-0.025}^{+0.017}$&	
$0.019_{-0.025}^{+0.035}$&
$-$
\\
SALT+LS&
$0.1552_{-0.0004}^{+0.0003}$& 
$2.521_{-0.037}^{+0.037}$&	
$0.0829_{-0.0089}^{+0.0088}$&
$-0.0546_{-0.0082}^{+0.0082}$  & 
$-$&
$-$ &
$0.0861_{-0.0047}^{+0.0049}$&	
$0.033_{-0.0028}^{+0.0029}$&
$-$
\\
LS2&
$0.18_{-0.01}^{+0.01}$& 
$2.00_{-0.50}^{+0.50}$&	
$-0.41_{-0.07}^{+0.11}$&
$-0.33_{-0.06}^{+0.09}$  & 
$-0.01_{-0.06}^{+0.06}$&
$0.02_{-0.00}^{+0.00}$ &
$-0.13_{-0.04}^{+0.04}$&	
$-0.10_{-0.04}^{+0.04}$&
$-$
\\
\enddata

\end{deluxetable*}

Our best-fit parameters, as shown in Table~\ref{tab:snz-best-fit}, differ significantly from those reported by G23 and P23. Previous models suggested that \snz is an Einstein cross system with a relatively low ellipticity (e.g. G23, $q = 0.65$), and with the lensed images located relatively far away from the critical curve. In contrast, our analysis reveals a completely different configuration, with significant implications for the interpretation of the system.

Our model predicts a higher ellipticity, with an axis ratio of $q = 0.379$, in sharp contrast to most of the earlier results---although GLEE's model \citep{pierel2023a} also has a relatively high ellipticity ($q = 0.470$). This ellipticity alters the geometry of the lensing configuration, resulting in an asymmetric distribution of the lensed images and modifying their proximity to the critical curve. The mass profile slope \gma also differs from previous models. In P23, three of the five lens models provided predictions for \gma (Table~\ref{tab:snz-best-fit}). The LS2 and GLEE models, which excluded SN luminosity data (flux), predicted near-isothermal profiles with $\gma \approx 2.0$. The SALT+LS model, which incorporated SN luminosity, found a steeper slope of $\gma = 2.5$. Our best-fit model predicts an intermediate slope of $\gma = 2.11$, higher than isothermal but less steep than the SALT+LS result. This value arises from the inclusion of positional, flux, and time-delay observations. The two remaining models in P23 and the model in G23 utilized Singular Isothermal Ellipsoid (SIE) mass profiles and only positional data.

All models in P23 and G23 found the source to be located near the center of the caustic, resulting in lensed images that were relatively symmetric in position but significantly distant from the critical curve. This configuration fails to explain the observed brightness asymmetry, where images A and C are considerably brighter than B and D, despite their nearly symmetric spatial arrangement (see Figure~\ref{fig:snz-obs}). This discrepancy raises this question: why do the brightnesses of the images differ so markedly when the images are fairly symmetrical relative to the critical curve?

Our model addresses this tension by predicting a quite unexpected lensing geometry. The combination of a slightly-higher-than-isothermal slope (\(\gamma = 2.11\)) and high ellipticity produces a critical curve with a distinct ``round-cornered astroid" shape\footnote{Such a shape was also found in \citet{riordan2024} when including a multipole expansion. In Section~\ref{sec:development-knots}, we show that a sufficiently high ellipticity, in combination with a steep mass slope can also reproduce this shape.}. This geometry places images A and C almost directly on the critical curve, resulting in their higher brightness, while images B and D are located somewhat further away from the critical curve, appearing dimmer. 
That is, in previous models, the four images are more or less ``equivalent'' in terms of their distance from the critical curve, resulting in similar predicted brightness, which is in tension with the observations. Our best-fit model predicts A and C to be brighter than B and D, which conforms to the observations.
Thus, this configuration successfully reproduces the observed brightness asymmetry without invoking additional effects such as microlensing or differential extinction for 
three
of the four images.
Simultaneously, the model successfully predicts all image positions. Notably, this outcome contrasts with earlier models that struggled to reconcile the image positions and fluxes using a smooth lens model. For instance, the SALT+LS model in P23 attempted to fit both positional and flux constraints. Their best compromise required a higher than usual but not unphysical value $\gma = 2.5$ and a slightly smaller Einstein radius and still, they failed to predict fluxes and positions as accurately as our best-fit model. Similar tensions were identified in the GLEE model, as shown in P23 (Figure~12).

\sbj{The final model presented in P23 (constructed as the average of all models that used positional data only) successfully recovered the positions of the four images and the flux ratios for two of them. 
However, all models in P23 predicted image arrival-time orderings that differ markedly from those inferred by our model. For instance, in Figure~\ref{fig:snz-best-model}, image~A is the trailing image in P23's model, while it becomes the leading image in our model; conversely, P23 predicts image~B to arrive first.
For systems with a sufficiently large angular separation---and thus appreciable time delays---this difference would yield a decisive preference for the correct model.
Still, the refined measurements from L24 indicate that image~A indeed arrived first, although the improved uncertainties remain large enough to allow images~B or D to be the leading images. This revised ordering is fully consistent with our model, predicting image~A to arrive earliest and thus to have positive parity---a result that follows naturally from the lensing geometry we found.
Moreover, two additional measurements from L24 shift in the direction favored by our model: image~B exhibits an even larger observed time delay relative to image~A, and its negative uncertainty is reduced.  
Taken together, these updates make our model more consistent with the current observations than the previous models.}

\begin{minipage}{\linewidth}
\vspace{5pt}
\makebox[\linewidth]{
  \includegraphics[keepaspectratio=true,scale=0.5]{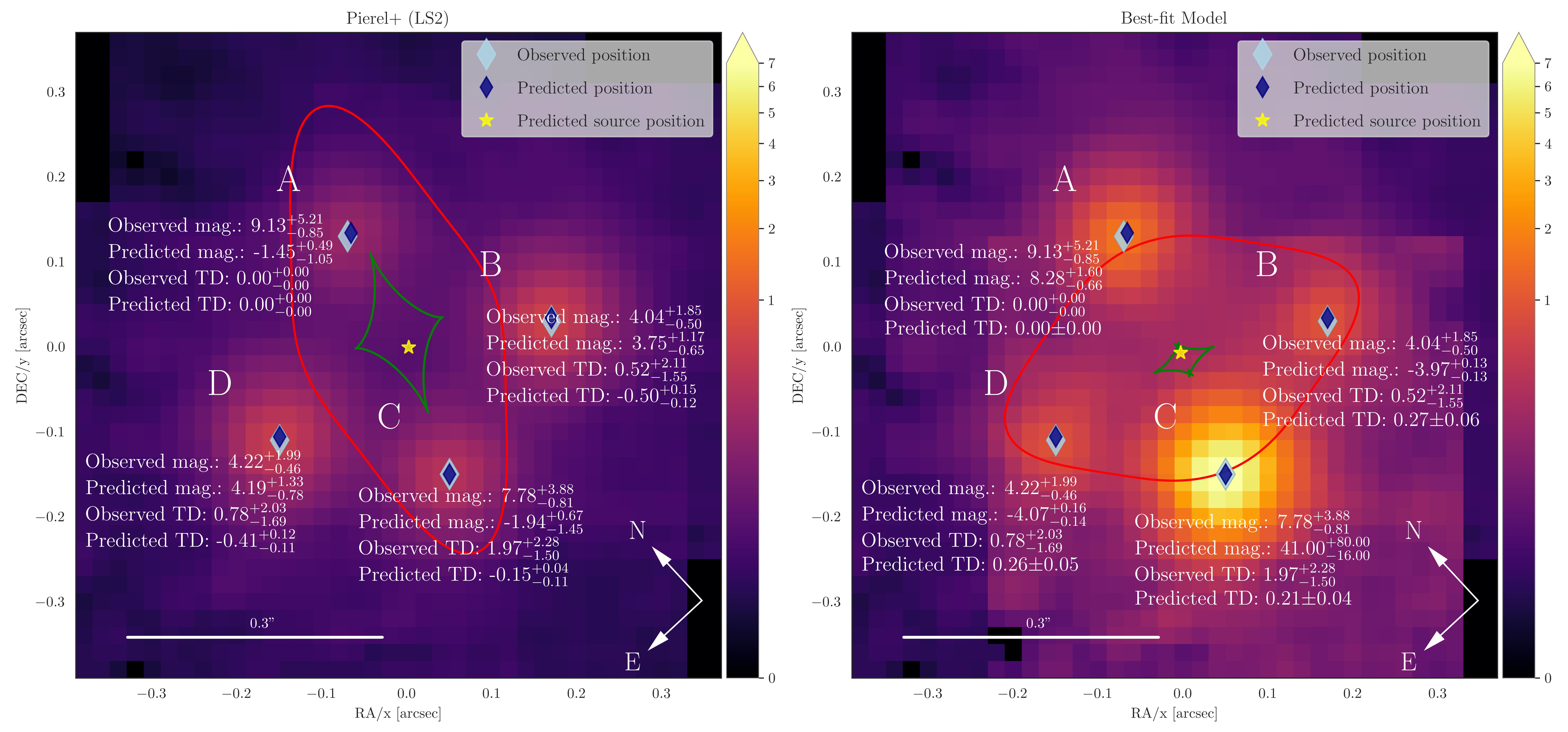}
  }
\captionof{figure}{Comparison of best-fit models for \snz: the left panel shows P23 (LS2) model, whereas the right panel shows our best-fit model. For each image, observed and predicted time delay and magnification are reported. Observed positions are shown in light blue and predicted positions in dark blue. The critical curve is drawn in red while in green is the caustic. The predicted source position is shown as a yellow star. The orientation and labeling match those of Figure~\ref{fig:snz-obs}.}
\label{fig:snz-best-model}
\vspace{10pt}
\end{minipage}

\begin{figure}[H]
  \centering
  \vspace{0.2cm}
  \includegraphics[scale = 0.33]{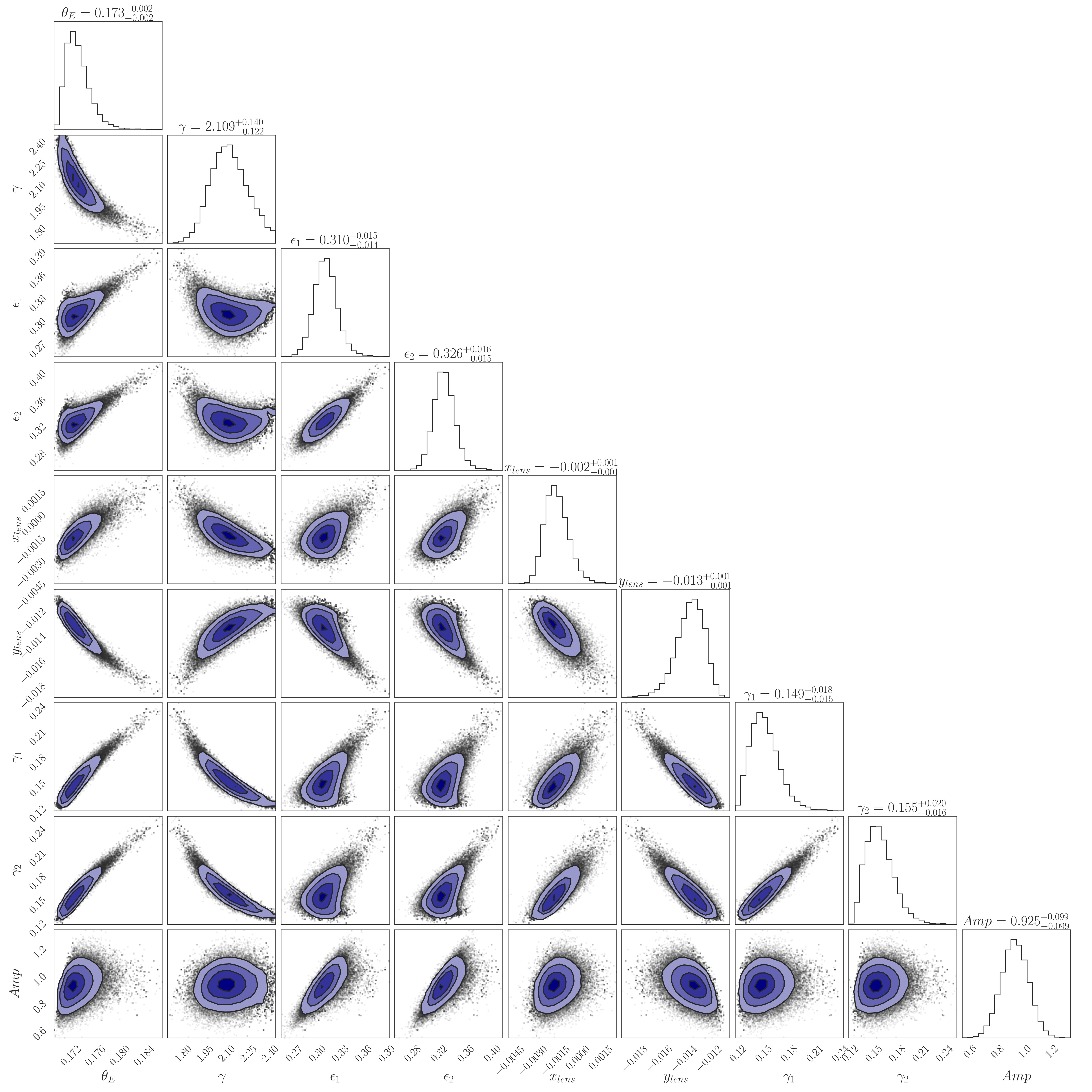}
  \begin{tikzpicture}[remember picture,overlay]
      \node at (-4.0cm,15.0cm) 
      {\includegraphics[scale = 0.40]{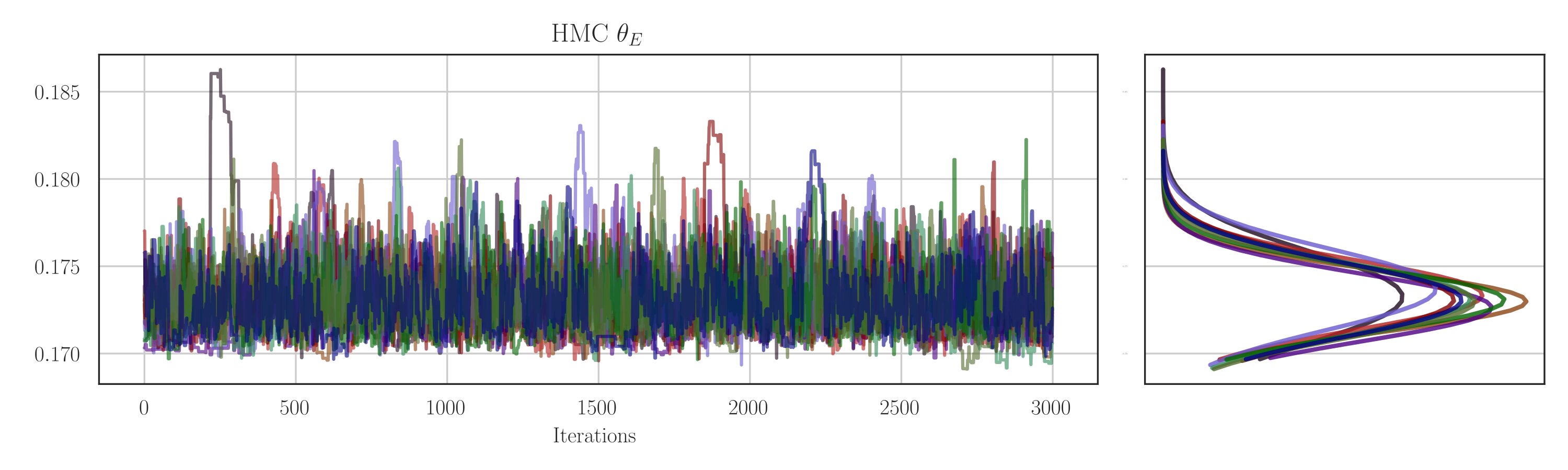}};
    \end{tikzpicture}
  \begin{tikzpicture}[remember picture,overlay]
      \node at (-4.1cm,12.0cm) 
      {\includegraphics[scale = 0.40]{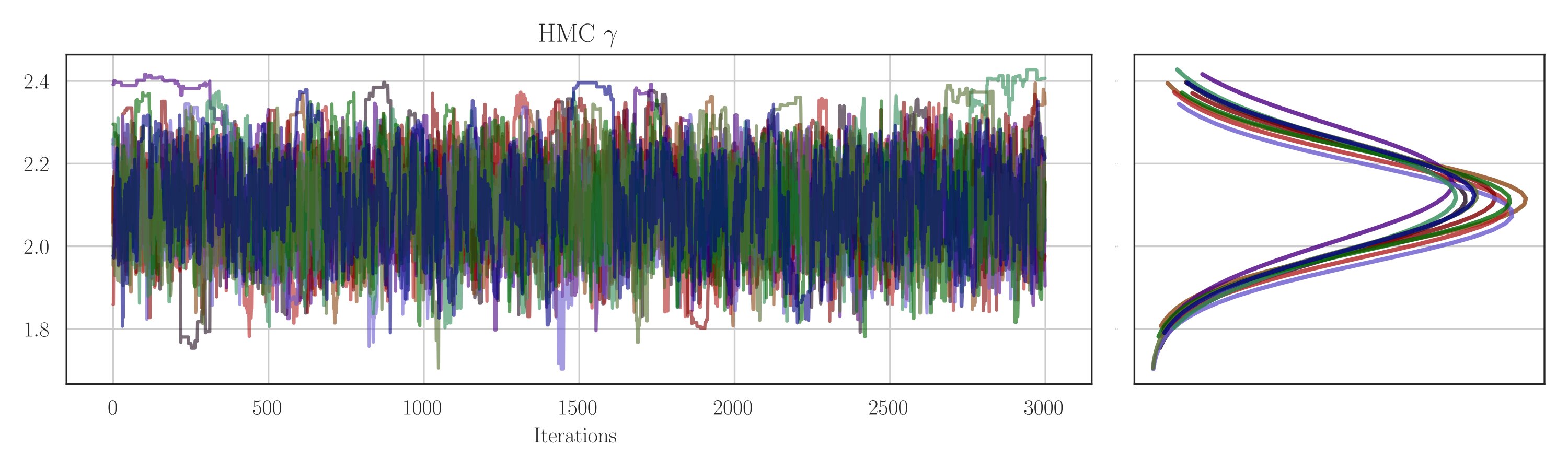}};
    \end{tikzpicture}
  \caption{\snz. The corner plot shows our sampling results for every parameter in our model, achieving $\rhat_{max} = 1.004$, and ESS $= 2989-37869$ within 6 min.\ and 35 sec.\ of modeling time. In the inset, ten chains of HMC demonstrate statistical consistency within the sampling of the Einstein radius, $\theta_E$, and the mass slope, \gma.}
  \label{fig:snz-corner}
\end{figure}

\xh{Previous models reported flux anomalies and were unable to correctly predict flux and positions simultaneously.} For our model, Figure~\ref{fig:snz-best-model} shows observed and predicted values for position, flux and time-delay along with our best-fit model. The recovered positions are in excellent agreement with the observations.
\textcolor{black}{Furthermore, our lens model reproduces the magnifications of images A, B, and D very well. Regarding image C, in our model, it lies very close to the critical curve, making its observed brightness extremely sensitive to even small deviations between the true mass distribution and our assumed parametric form. The magnification of such near-critical images is also highly sensitive to the exact source position, an uncertainty already reflected in the predicted brightness interval that encloses the observed value. These effects naturally account for the brightnesses of all four images without requiring additional components in the mass model, including microlensing.}
Hence, we deem our model as a more plausible explanation for the observations.
In Figure~\ref{fig:snz-corner} a full cornerplot is shown and, for our best-fit model, the \rhat values for all parameters are smaller than $1.004$. Our model is also very fast, with a total modeling time of 6 minutes and 35 seconds.
These results demonstrate, in a real system, that not only can \gma be constrained  using point source information alone,
but that the standardizablity of Type~Ia SNe can be used to constrain \gma in a unique way---through the flux term of our loss function (Equation~\ref{eqn:flux-loss}).

%% file: discussion.tex
\subsection{SN iPTF16geu and Mass Sheet Degeneracy}\label{sec:16geu-disc}
A uniform mass concentration along the line-of-sight can modify the convergence map of the mass distribution without altering the observed lensing configuration \citep{Schneider2013a}. To address this degeneracy in lens models, M20 conducted an analysis integrating velocity dispersion measurements with their lens model. Their findings indicated a mass slope, $\gamma$, of 1.8.
The presence of a mass sheet of this type would effectively reduce the $\gamma$ value inferred from lens modeling alone. Alternatively, when using lens modeling to predict velocity dispersion, the resulting prediction would exceed the observed velocity dispersion \citep[e.g.][]{Sahu2024a}. 
The best-fit mass slope from our model is consistent with $\gamma = 1.8$ within $1\sigma$, suggesting that the mass-sheet is unlikely a significant factor of influence for this system.

\subsection{SN~Zwicky}\label{sec:snzwicky-disc}

\subsubsection{Image Symmetry, Parity, and the Critical Curve}\label{sec:sym}



\begin{figure}[H]
    \centering
    \includegraphics[width=1.0\linewidth]{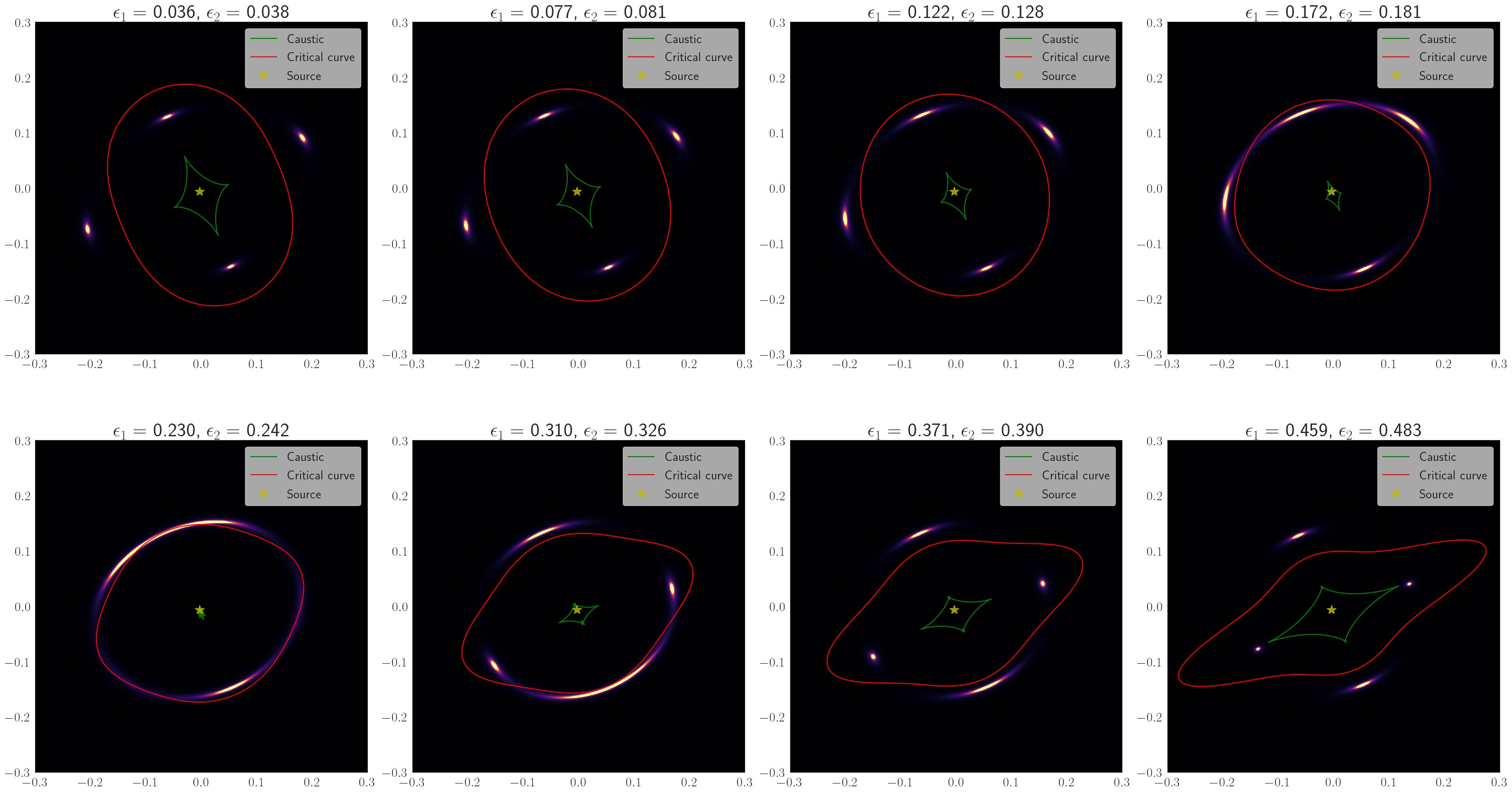}
    \caption{\sbj{In order to show the intricate structure of the caustic, we plot high-resolution simulations (pixel size~$= 0.002''$) generated using the best-fit parameters of \snz (Table~\ref{tab:snz-best-fit}), with $\gma = 2.11$. Only the axis-ratio is allowed to vary over the range $0.9 \geq q \geq 0.2$, in the form of the eccentricities $\epsilon_1, \epsilon_2$ (see Eq.~\ref{eqn:e1,e2-vs-q-phi}).} The best-fit model for \snz corresponds to $\epsilon_1 = 0.310, \epsilon_2 = 0.326$.}
    \label{fig:knots-ecc}
\end{figure}

Figure~\ref{fig:knots-ecc} illustrates the results of eight simulations exploring the effects of eccentricity (equivalently axis-ratio $q$). As the eccentricity increases, the system becomes more elongated.
At $q = 0.5$, with $\epsilon_1 = 0.230$ and $\epsilon_2 = 0.242$,
the resulting lensing configuration places two images very close to the critical curve (images A and C for \snz), with the other two located farther away (images B and D).
That is, although the \snz images lie in relatively symmetric positions, the orientation of the elongated critical curve in our best‐fit model 
accounts for the observed brightness asymmetry.

It is clear that the models from P23 could be readily distinguished from our best-fit model if the lensed host-galaxy light could be robustly modeled. However, in this system, the combination of the SN lying far from the host-galaxy center and the small Einstein radius results in insufficient detectable host emission to constrain the model. While jointly modeling the point source and the lensed host galaxy would be ideal (and would be straightforward to implement, which we plan to pursue in future work), we note that in both lensed quasar systems and lensed SNe the host galaxy being considerably fainter than the point source is not uncommon, highlighting the importance of the pipeline presented here.

We show in this work that when the time delays are sufficiently large, modeling the lensed point source alone is also capable of distinguishing the P23 models from ours. In particular, we demonstrate the constraining power of the information carried by the lensed point source---namely, image positions, fluxes (for SNe~Ia), and time delays.
For quasars and non-Ia SNe, our current pipeline can be extended by incorporating flux ratios (Ratier-Werbin et al., in prep.).

\subsubsection{Caustic and Critical Curve Structures}\label{sec:development-knots}

\begin{figure}[H]
    \centering
    \includegraphics[width=1.0\linewidth]{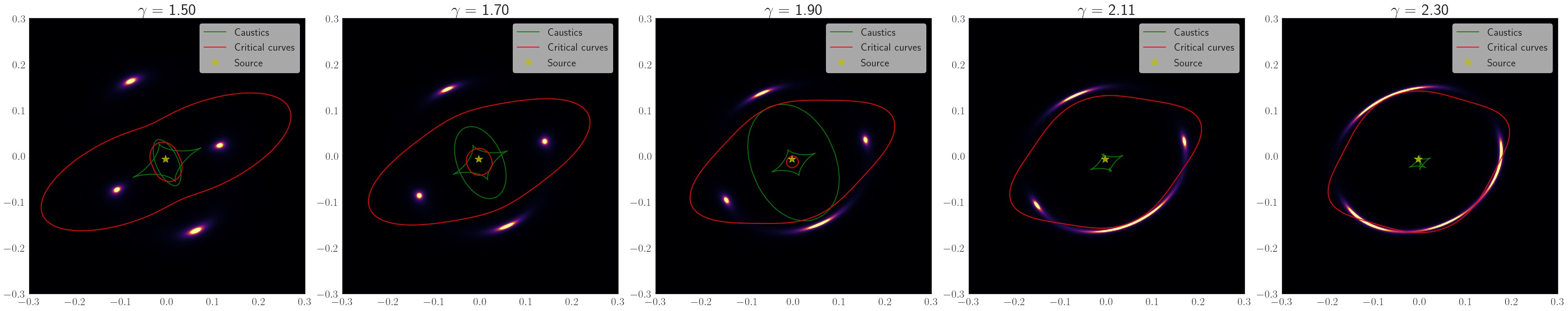}
    \caption{\sbj{High-resolution simulations (pixel size~$= 0.002''$) generated using the best-fit parameters of \snz (Table~\ref{tab:snz-best-fit}), varying only the mass slope over the range $1.5\leq \gma \leq 2.3$}. The best-fit model for \snz corresponds to $\gma = 2.11$.}
    \label{fig:knots-gamma}
\end{figure}

\citet[][Figure~2]{riordan2024} showed that fourth order angular Fourier perturbations on SIE create ``knots'' in the caustic.
We report that in a certain regime of the parameter space of EPL such knots can appear without the introduction of angular Fourier perturbations.
In Figure~\ref{fig:knots-ecc} we show that for the EPL model, knots can appear in the caustic as the eccentricity becomes sufficiently high.
In addition, Figure~\ref{fig:knots-gamma}
\xh{demonstrates that increasing the mass slope \gma can also cause the knots to appear in the caustic.}
\sbj{Notably, even for the relatively low value of $\gamma = 2.11$---\snz's best-fit mass slope---}such caustic structure can still arise.


 
Finally, as is well-known, for $\gma < 2$, an inner caustic and critical curve form. Figure~\ref{fig:knots-gamma} shows that at $\gma = 1.5$, the inner caustic is nearly entirely contained within the inner critical curve, unlike the cases with higher mass slopes where the opposite is true. \sbj{In fact, in a system with axis-ratio $q = 1.0$, i.e. no ellipticities according to Eq.~\ref{eqn:e1,e2-vs-q-phi}}, the inner critical curve and caustic precisely coincide when $\gma = 1.5$.
To our knowledge, this specific property has not been previously reported in the literature.

\subsection{Convergence Criteria and the \rhat Threshold}

All systems presented in this paper satisfy the commonly adopted criterion $\rhat < 1.1$ and exhibit sufficiently large effective sample sizes.
While $\rhat < 1.1$ remains the standard threshold in cosmological inference,
\citet{vehtari2021a} argued that $\rhat < 1.01$ provides a more reliable indication of convergence.
For four out of the five simulated archetypal configurations, with full forward modeling, we reach $\rhat < 1.01$ for all parameters; with the doubly-lensed system converging to $\rhat = 1.015$. 
Importantly, we have achieved $\rhat < 1.01$ for an additional simulated system that directly constrains \ho. 
\emph{For the two real systems, both reach $\rhat < 1.01$.}
Reaching $\rhat < 1.01$ typically requires substantially longer runtimes than to reach $\rhat < 1.1$, by roughly an order of magnitude---from seconds to minutes.
We note that the systems modeled in this work involve nine model parameters, and achieving similarly stringent convergence may be more challenging for higher-dimensional parameter spaces. 
Our results nevertheless demonstrate that such stringent convergence is feasible for strong lensing, and with continued improvements in computational resources---such as more powerful GPUs---this threshold may become increasingly attainable for a broader range of strong-lensing applications.


%% file: conclusion.tex
In this paper, we introduce a novel method for modeling strongly lensed point sources, 
applicable to \Iae and quasars. We demonstrate its capabilities by applying it to six simulated systems. Our findings affirm that the discovery and observation of lensed \Iae with Einstein radii $\gtrsim 1''$ are of great value for determining \ho.
Our models of such simulated systems reveal that even with 
full forward modeling, which yields broader but more reliable uncertainties, a single system can still provide a precise constraint on \ho.

We also apply our modeling approach to two observed systems: \geu and \snz. For \geu, our results are within $1\sigma$ agreement with previous models \citep{Mortsell2020a}. We conclude that a smooth model alone is insufficient to account for the observations, and agree that microlensing is a plausible explanation. In the case of \snz, our model provides an alternative to previously reported models \citep{goobar2023a, pierel2023a}.
Our best-fit model successfully predicts the positions for all four images \emph{and} magnifications for three images within model uncertainties. 
Due to our treatment of the magnification singularity, our pipeline can robustly analyze images near the critical curve and simultaneously incorporate position, flux, and time-delay constraints in the loss function, providing a complete description of lensed SN~Ia observables.

Our sampling results demonstrate high performance: for all parameters, including intrinsic SN brightness and \ho, we achieve a potential scale reduction factor, $\hat{R}$, below 1.01 and an effective sample size (ESS) of $\gtrsim \mathcal{O}(10^3)$ for all systems---observed and simulated---except the double configuration. The average modeling time for our simulated systems is 13 minutes and 49 seconds on four A100 GPUs. 

In the preparation of this manuscript, a new galaxy-scale lensed SN was discovered, SN~2025wny \citep{johansson2025, taubenberger2025}. We are in the process of modeling this system and will report the results in a future publication.

Overall, this work demonstrates the feasibility of rapid and statistically rigorous modeling of strongly lensed point sources. While this study focuses on lensed supernovae—particularly SNe~Ia—the framework presented here can be readily extended to lensed quasars and non-Ia supernovae by incorporating flux ratios. Moreover, combining this point-source modeling framework with host-galaxy modeling using the original pixel-based \gigal approach offers a clear path toward even tighter constraints on \ho.


